\documentclass[english]{article}
\usepackage[utf8]{inputenc}
\usepackage[margin=2.5cm]{geometry}
\usepackage{babel}

\usepackage{alias}
\usepackage[utf8]{inputenc}
\usepackage{graphicx}
\usepackage{natbib}
\usepackage{amsmath,amsfonts,amssymb,amsthm}
\usepackage{algorithmic, algorithm}
\usepackage{physics}
\usepackage{wrapfig}

\usepackage{cleveref}
\usepackage{float}
\usepackage{url}

\newcommand{\X}{\mathbb{R}^d}
\newcommand{\bN}{\mathbb N}

\newcommand{\ssp}[1]{\langle #1 \rangle}
\newcommand{\cM}{\mathcal M}
\newcommand{\cC}{\mathcal C}
\newcommand{\cG}{\mathcal G}

\newcommand{\diff}{\mathrm{d}}

\newcommand{\Id}{\mathop{\mathrm{Id}}\nolimits}

\newcommand{\FI}{\mathrm{I}}

\newcommand{\cP}{\mathcal{P}}
\newcommand{\KL}{\mathop{\mathrm{KL}}\nolimits}

\DeclareMathOperator*{\argmin}{argmin}

\newcommand{\ess}{\textrm{ESS}}
\DeclareMathOperator*{\dom}{dom}

\theoremstyle{plain}

\usepackage{authblk}

\title{A connection between Tempering and Entropic Mirror Descent}
\author[1]{Nicolas Chopin}
\author[1]{Francesca R. Crucinio\thanks{email: francesca\_romana.crucinio@kcl.ac.uk}}
\author[1]{Anna Korba\thanks{email: anna.korba@ensae.fr}}
\affil[1]{ENSAE, CREST, IP Paris}

\date{ }
\begin{document}
\maketitle

\begin{abstract}
    This paper explores the connections between tempering (for Sequential Monte Carlo; SMC) and entropic mirror descent to sample from a target probability distribution whose unnormalized density is known.
    We establish that tempering SMC 
    corresponds to 
    entropic mirror descent applied to the reverse Kullback-Leibler (KL) divergence and obtain convergence rates for the tempering iterates.
    Our result motivates the tempering iterates from an optimization point of view, showing that tempering can be 
    seen as a descent scheme of the KL divergence with respect to the Fisher-Rao geometry, in contrast to Langevin dynamics that perform descent of the KL with respect to the Wasserstein-2 geometry.
    We exploit the connection between tempering and mirror descent iterates to justify common practices in SMC and derive adaptive tempering rules that improve over other alternative benchmarks 
    in the literature.
\end{abstract}

\section{Introduction}

Sampling from a target probability distribution whose density is known up to a normalization constant is a fundamental task in computational statistics and machine learning. It can be naturally formulated as optimizing a functional measuring the dissimilarity to the target probability distribution, typically the Kullback-Leibler (KL) divergence. From there, it is natural to consider optimization schemes over the space of probability distributions, to design a sequence of distributions approximating the target one. Depending on the chosen geometry over the search space and the time discretization, one may obtain different schemes.

For instance, one possible framework is to restrict the search space to the Wasserstein space, i.e. probability distributions with bounded second moments equipped with the Wasserstein-2 distance \citep{ambrosio2008gradient}. The latter is equipped with a rich Riemannian structure \citep{otto2000generalization}, which makes it possible to define Wasserstein-2 gradient flows, i.e. paths of distributions decreasing the objective functional of steepest descent according to this metric. It is well-known that the Wasserstein gradient flow of the KL can be implemented by a Langevin diffusion on the ambient space \citep{jordan1998variational} and easily discretized in time, resulting for instance in the Langevin Monte Carlo (or Unadjusted Langevin) algorithm \citep{roberts1996exponential}. The latter is one of the most famous Markov Chain Monte Carlo (MCMC) algorithms - maybe the most canonical - that generate Markov chains in the ambient space, whose law approximates the target distribution for a large time horizon. 
It is known to converge fast when the target distribution has a smooth and strongly convex potential \cite{durmus2019analysis}, or is satisfies a relaxed log-Sobolev assumption \cite{vempala2019rapid}.
Alternative time discretizations of the KL Wasserstein gradient flow \citep{salim2020wasserstein, mou2021high} or its gradient flow with respect to similar optimal transport geometries have been considered in the literature to propose alternative algorithms
\citep{liu2017stein,garbuno2020interacting}, but their convergence also depends strongly on the log-concavity of the target.

Another possible framework is to cast the space of probability distributions as a subset of a normed space of measures (such as $L^2$), and to consider the duality of measures with continuous functions and the mirror descent algorithm that relies on Bregman divergences geometry, as recently considered in \citet{ying2020mirror,chizat2021convergence,aubin2022mirror}. While both frameworks yield optimization algorithms on measure spaces, the geometries and algorithms are very different (in particular notions of gradients and convexity). Mirror descent produces multiplicative ("vertical") updates on measures allowing for change of mass, while Wasserstein flows corresponds to displacement of fixed mass particles supporting the measures ("horizontal" updates). 
Moreover, as recently highlighted in \cite{aubin2022mirror},  the (reverse) KL as an objective loss for sampling is actually strongly convex \emph{whatever the target $\pi$} and smooth in a mirror descent geometry induced by the KL as a Bregman divergence. 
In contrast, it is known that the KL as an optimization objective is not smooth in the Wasserstein geometry \citep{wibisono2018sampling}, and as we have said earlier it enjoys convexity properties only if the target distributions does as well. 
Above all,  the latter scheme, namely \emph{entropic mirror descent} 
on the KL yields a sequence of distributions that takes the simple form of a geometric mixture between an initial  distribution and the target, a well-known sequence referred to as \emph{tempering} (or \emph{annealing}) in the Monte Carlo literature \citep{neal2001annealed}. 
Interestingly, entropic mirror descent on an objective functional can be seen as an Euler discretization of its Fisher Rao gradient flow \cite{domingo2023explicit}.

Algorithms approximating the tempering sequence offer an alternative to Langevin-based MCMC methods, and are often employed when the latter suffer from poor mixing \citep{syed2022non} or when estimates of the normalizing constant are needed \citep{gelman1998simulating}.
A number of algorithms have been proposed to approximate the tempering sequence, including sequential Monte Carlo (SMC; \cite{del2006sequential}), annealed importance sampling (AIS; i.e.  an SMC sampler in which no resampling occurs \cite{neal2001annealed}), and parallel tempering (PT; \cite{geyer1991markov}). Independently, a number of schemes aiming at directly approximating the entropic mirror descent iterates on the KL have also been proposed \citep{dai2016provable, korba2022adaptive}.

Choosing the right scheduling of temperatures for the sequence of tempered targets (or equivalently the step-sizes as we will explain in more detail in this paper), is critical in practice. 
Adaptive selection of the sequence of temperatures is an active area of research in the AIS literature; however, many of this strategies are intractable \cite{gelman1998simulating}, costly \cite{kiwaki2015variational}, limited to exponential families \cite{grosse2013annealing},
or numerically unstable \cite{goshtasbpour2023adaptive} 
as we show in our experiments. 
In the SMC literature, the sequence of temperatures is normally chosen adaptively using the effective sample size, a proxy for the variance of the importance sampling weights \citep{jasra2011inference}.
Adaptive strategies are widely used in practice but theoretical studies on how to select the tempering iterates are limited to specific target distributions (see \cite{beskos2014stability} for i.i.d. targets and \citet[Proposition 17.2]{chopin2020introduction}, \citet[Section 3.3]{dai2022invitation} for Gaussian targets). 

In this paper, we investigate the links between tempering and mirror descent and show that algorithms which sample from the tempering sequence (such as SMC) can be seen as numerical approximations to entropic mirror descent applied to the KL divergence, i.e. a time-discretization of the KL gradient flow in the Fisher Rao geometry.
We thus establish a parallel result to that of \cite{jordan1998variational,wibisono2018sampling} which shows that algorithms based on the Langevin diffusion can be seen as numerical approximations of gradient flow of the KL in the Wasserstein-2 geometry.

We adapt the proof of convergence of mirror descent in the space of measures of \citet[Theorem 4]{aubin2022mirror} to the case of varying step sizes and obtain a convergence rate for the tempering iterates. From this optimization point of view, we also justify the popular adaptive strategy that identifies the tempering sequence by ensuring that the (KL, Bregman) divergence between two consecutive distributions in the tempered sequence is small and constant. We show that for a generic target distribution, this tempering sequence obeys a differential equation, that can be solved easily analytically in some simple cases that we highlight, or by a simple numerical approximation based on particles in general cases. 


The paper is organized as follows. \Cref{sec:background} provides the relevant background on mirror descent on the space of measures. \Cref{sec:tempering} details the connection between tempering and entropic mirror descent and its consequence on designing tempering schedules. \Cref{sec:algorithms} discusses different strategies that were employed in the literature to approximate entropic mirror descent and their pros and cons. In \Cref{sec:relevant_work} we connect our results with relevant works in the SMC/AIS literature.

\section{Mirror descent on measures }\label{sec:background}

In this section, we recall the main steps to derive the mirror descent algorithm on the space of measures. The reader may refer to \citet{aubin2022mirror} for a detailed introduction.

\textbf{Notations.} Fix a vector space of (signed) measures $\cM(\X)$. 
Let $\cM^*(\X)$ the dual of $\cM(\X)$.
    For $\mu\in \cM(\X)$ and $f\in \cM^*(\X)$, we denote $\ssp{f, \mu}=\int_{\X}f(x)\mu(dx)$. 
We denote by $\cP(\R^d)$ the set of probability measures on $\R^d$. The Kullback-Leibler divergence is defined as follows: for $\nu,\mu \in \cP(\R^d)$, $\KL(\nu|\mu)=\int \log(\nicefrac{d\nu}{d\mu}) d\nu$ if $\nu$ is absolutely continuous w.r.t. $\mu$ with Radon-Nikodym density $\nicefrac{d\nu}{d\mu}$, and $+\infty$ else.

\subsection{Background on Mirror Descent}

Let $\cF:\cP(\X)\rightarrow \R^+$ be a functional on $\cP(\X)$. Consider the optimization problem
\begin{equation*}
    \min_{\mu \in \cP(\R^d)} \cF(\mu).
\end{equation*}
Mirror descent is a first-order optimization scheme relying on the knowledge of the derivatives of the objective functional, and a geometry on the search space induced by Bregman divergences. These two notions are introduced in the following definitions. 
\begin{definition}\label{def:first_var}
If it exists, the \textit{first variation of $\cF$ at $\nu$} is the  function $\nabla \cF(\nu):\X \rightarrow \R$ s. t. for any $\mu \in \mathcal{P}(\X)$, with $\xi = \mu-\nu$:
\begin{equation}
\label{eq:first_var}
\lim_{\epsilon \rightarrow 0}\nicefrac{1}{\epsilon}(\cF(\nu+\epsilon  \xi) -\cF(\nu))
=\ssp{\nabla\cF(\nu), \xi}
\end{equation} 
and is defined uniquely up to an additive constant.
\end{definition}

\begin{definition}
Let $\phi:\cP(\X)\rightarrow \R^+$ a convex functional on $\cP(\X)$. The $\phi$-Bregman divergence is defined for any $\nu,\mu \in \cP(\X)$ by:
\begin{equation}
\label{eq:breg_probas}
     B_{\phi}(\nu|\mu)=\phi(\nu)-\phi(\mu)-\ssp{\nabla \phi(\mu), \nu- \mu}
\end{equation}
where $\nabla \phi(\mu)$ is the first variation of $\phi$ at $\mu$.
\end{definition}

Consider a sequence of step-sizes $(\gamma_n)_{n\ge 0}$. Starting from an initial $\mu_0\in \cP(\R^d)$, one can generate a sequence $(\mu_{n})_{n \in \bN}$  
\begin{align}\label{eq:prox_algo}
    \mu_{n+1}=\argmin_{\mu \in \cP(\X)}\left\{ \cF(\mu_n) +\ssp{ \nabla \cF(\mu_n), \mu-\mu_n}+ (\gamma_{n+1})^{-1} B_{\phi}(\mu|\mu_n) \right\}.
\end{align}
The first variation of $\phi$, denoted $\nabla \phi:\cP(\X)\rightarrow \cC(\X)$, maps an element of the primal (a distribution) to an element of the dual (a function). In particular, writing the first order conditions of \eqref{eq:prox_algo} we obtain the dual iteration
\begin{equation}   \label{eq:dual_iter} \nabla \phi(\mu_{n+1} ) -  \nabla \phi(\mu_n) = - \gamma_{n+1} \nabla \cF(\mu_n).
\end{equation}
The scheme \eqref{eq:prox_algo}-\eqref{eq:dual_iter} is referred to as mirror descent \citep{beck2003mirror}. It has been shown recently in \cite{aubin2022mirror}, that the mirror descent scheme converges linearly as soon as there exists $0\le l \le L$ such that $\cF$ is relatively $l$-strongly convex and $L$-smooth with respect to $\phi$, a condition that can be written as $l B_{\phi}(\nu|\mu)\le B_{\cF}(\nu|\mu)\le L B_{\phi}(\nu|\mu)$, for constant stepsizes smaller than $\nicefrac{1}{L}$; extending the results of \cite{lu2018relatively} to the infinite-dimensional setting of optimization over measures. In particular it applies to the case where both the objective and Bregman divergence are chosen as the KL.

\subsection{Entropic mirror descent on the KL}
Consider the negative entropy functional:
\begin{equation}\label{eq:entropy}
    \phi: \mu\mapsto \int \log (\mu(x))d\mu(x)
\end{equation}
where $\mu$ also denotes its density w.r.t. the Lebesgue measure on $\R^d$. Since the  first variation of $\phi$ at $\mu$ writes $ \nabla \phi(\mu)= \log(\mu)$,
 one gets from \eqref{eq:breg_probas} that $B_{\phi}(\nu|\mu)=\KL(\nu|\mu)$, 
and choosing $\phi$ as $\eqref{eq:entropy}$ yields the following multiplicative update named \emph{entropic mirror descent}:
\begin{equation}
\label{eq:md}
    \mu_{n+1} \propto \mu_n e^{-\gamma_{n+1}\nabla \cF(\mu_n)}
\end{equation}
by exponentiating \eqref{eq:dual_iter}. Notice that the latter scheme is an Euler discretization of the Fisher-Rao gradient flow of the functional $\cF$, as noticed in \cite{domingo2023explicit} (see \Cref{sec:fisher_rao} for details).

Moreover, if $\cF(\mu)=\KL(\mu|\pi)$ (the reverse KL with respect to $\pi$), $\nabla \cF(\mu)=\log(\nicefrac{\mu}{\pi})$ and we obtain \emph{entropic mirror descent} on the KL iterates:
\begin{equation}\label{eq:md_temp}
    \mu_{n+1}\propto\mu_n^{(1-\gamma_{n+1})}\pi^{\gamma_{n+1}}.
\end{equation}
Since $\cF$ is 1-strongly convex and 1-smooth with respect to ${\phi}$ since $B_{\cF}=B_{\phi}$ (i.e. $l=L=1$), as soon as one uses step-sizes $\gamma_n <1$, the KL objective decreases at each step of the scheme \eqref{eq:md_temp}, and converges at a linear rate as stated in the Proposition below.
\begin{proposition}\label{prop:rate_md} Let $\mu_0 \in \cP(\R^d)$ an initial distribution. Entropic mirror descent iterates on  $\cF=\KL(\cdot|\pi)$ as defined in \eqref{eq:md_temp} converge at a rate:
\begin{equation}\label{eq:rate_md}
    \KL(\mu_n|\pi) \le \frac{C_n}{\gamma_1} \KL(\pi|\mu_0);\; C_n^{-1} = \sum_{k=1}^n\prod_{i=1}^k\frac{\gamma_k/\gamma_1}{1-\gamma_i }.
\end{equation}
where $(\gamma_k)_{k=1}^{n}$ is the sequence of step-sizes. 
In particular, a simple induction argument shows that $C_n\leq \prod_{k=1}^n(1-\gamma_k)\to 0$ as $n\to \infty$ when $\gamma_n\leq 1$ for all $n\geq 1$. 
Hence, the mirror descent iterates~\eqref{eq:md_temp} satisfy
\begin{equation}\label{eq:rate_KL_upper_bound}
    \KL(\mu_n|\pi) \le (\gamma_1)^{-1} \prod_{k=1}^n(1-\gamma_k) \KL(\pi|\mu_0).
\end{equation}
\end{proposition}
The proof of \Cref{prop:rate_md} is given in \Cref{sec:proof_th_rate}. It extends the result of \citep[Theorem 4]{aubin2022mirror}, that could not be applied to varying step-sizes while it is the case for ours. We derived our result by carefully adapting the proof of \citep[Theorem 4]{aubin2022mirror} or \citep[Theorem 3.1]{lu2018relatively}; the extension is non-trivial and involves verifying a recursion that is the same as the one in the latter references for constant or decreasing step sizes (as detailed in \Cref{sec:decreasing_or_constant_step_size}) or a different one for general step-sizes (as detailed in \Cref{sec:general_step_size}). We consider the two cases separately, as this allows us to obtain sharper rates.
%

\Cref{prop:rate_md} shows that if $\KL(\pi|\mu_0)<\infty$ and the step sizes are smaller than $1$ (the inverse of the smoothness constant $L=1$), $C_n^{-1}\to \infty$, and entropic mirror descent on the KL converges to the target distribution. 
We note that \citet[Lemma 2]{korba2022adaptive} show a similar result on the total variation\footnote{notice that Pinsker's inequality combined with our result \eqref{eq:rate_KL_upper_bound} recover their rate on the TV.}. We also note that our (discrete-time) rate is coherent with the convergence rate of its continuous-time counterpart, i.e. Fisher-Rao dynamics for the KL \citep[Theore 2.4]{lu2023birth}, that is known to converge exponentially fast under a warm-start assumption
on the support of the initial distribution with respect to the target. 
Finally, if the sequence of  $(\gamma_n)_{n\ge 0}$ is fixed to $\gamma$  constant, mirror descent converges at a linear rate proportional to $(1-\gamma)^{n}$, as already shown in \citet[Eq. (27)]{lu2018relatively}.

\section{A connection between Mirror Descent and Tempering}
\label{sec:tempering}
We now turn to the connection between entropic mirror descent and tempering, that, to the best of our knowledge, we are the first to to highlight and exploit (see \cite{domingo2023explicit} for a similar connection in continous time). In particular we will show that the tempering schedule is deeply connected to optimization/step-sizes dynamics of the corresponding entropic mirror descent scheme.

In the Monte Carlo literature, it is common to consider the following tempering (or annealing) sequence \citep{gelman1998simulating, neal2001annealed}
\begin{equation}\label{eq:tempering}
    \mu_{n+1}\propto\mu_0^{1-\lambda_{n+1}}\pi^{\lambda_{n+1}},
\end{equation}
where $0 = \lambda_0 < \lambda_1<\dots <\lambda_T = 1$, to sample from a target distribution $\pi$. There is a correspondence between \eqref{eq:tempering} and \eqref{eq:md_temp} if 
\begin{equation}\label{eq:connection_lambda_gamma}
    \lambda_n = 1- \prod_{k=1}^n(1-\gamma_k) 
\end{equation}
which by induction yields $\gamma_1 = \lambda_1$, $\gamma_n = (\lambda_n - \lambda_{n-1})/{1-\lambda_{n-1}}$ for $1\leq n < T$ and $\gamma_T=\lambda_T=1$. Notice that reversely, if we have a sequence $\gamma_n$ defined as $\gamma_n = (\lambda_n - \lambda_{n-1})/{1-\lambda_{n-1}}$, as soon as the $\lambda$'s are in $(0,1)$, $\gamma_n<1$, guaranteeing descent of the KL objective at each step.

In the tempering sequence~\eqref{eq:tempering}, $\lambda_T = 1$ to ensure that we are targeting the correct distribution $\pi$. In the case of the mirror descent iterates~\eqref{eq:md_temp} the convergence to $\pi$ is in the limit $n\to\infty$.
We can thus interpret~\eqref{eq:tempering} as performing $T-1$ mirror descent steps towards $\pi$ and then one final step bridging step to reach $\pi$. Hence, it is interesting to look at the speed of convergence of the iterates~\eqref{eq:md_temp} to gain some intuition on the number of bridging distributions $\mu_n$ necessary to get close enough to $\pi$ to guarantee that the final step (corresponding to $\lambda_T = 1$) is stable.
In this case, combining and \eqref{eq:rate_KL_upper_bound} and \eqref{eq:connection_lambda_gamma} we get that 
\begin{equation}\label{eq:rate_KL_upper_bound_lambda}
    \KL(\mu_n|\pi) \le 
   (\lambda_1)^{-1} (1-\lambda_n)
    \KL(\pi|\mu_0),
\end{equation}
which approaches 0 as $\lambda_n\to 1$
, and gives an explicit rate of convergence of the  sequence \eqref{eq:tempering}. 
Provided one can obtain an approximation of $\KL(\pi|\mu_0)$, 
we can infer the value of $\lambda_n$ necessary to guarantee that the $n-$th tempering iterate is sufficiently close to $\pi$.
Later in this section, we derive several examples.

\subsection{A principled strategy for tempering}\label{sec:strategy_tempered}

As the speed of convergence of the mirror descent iterates depends on the sequence $(\lambda_n
)_{n\geq 1}$, we now discuss relevant strategies to select temperatures, in the light of the optimization scheme. 

Notice that \eqref{eq:tempering} admits an exponential family representation \citep{brekelmans2020all, syed2021parallel}
\begin{align}
\label{eq:exp_fam}
    \mu_{n+1}(x)\equiv\mu_{\lambda_{n+1}}(x)\propto\mu_0 \exp\left\lbrace \lambda_{n+1}s(x)\right\rbrace
\end{align}
where $s(x):= \log \pi(x)/\mu_0(x)$. 

A popular strategy in the SMC/AIS literature to identify the sequence $(\lambda_n
)_{n=0}^T$ is to fix $\lambda_0 = 0$ and then select  $\lambda_n$ iteratively, ensuring that the $\chi^2$ divergence between successive distributions is constant and sufficiently small, e.g. setting $\chi^2(\mu_{n-1}|\mu_n)=\beta$ for some small value of $\beta$ (see \cite{jasra2011inference} for $\chi^2$ in SMC, and more recently \cite{goshtasbpour2023adaptive} for $\alpha$-divergences in AIS). This quantity is related to the variance of the importance weights and ensures that this variance remains low.

The following Proposition, whose proof can be found in Appendix \ref{sec:proof_fdiv}, shows that, up to higher order terms, the $\chi^2-$divergence can be replaced by any $f-$divergence whose $f$ is twice differentiable (see also \citet[Section 3.4]{amari2016information}), in particular the KL divergence. 
Let  $D_{f}(\lambda'|\lambda):=\int\mu_{\lambda}f(\mu_{\lambda'}/\mu_{\lambda})$
be the $f-$divergence of $\mu_{\lambda'}$ relative to $\mu_{\lambda}$.
\begin{proposition}
\label{prop:fdiv}
Provided $f$ is twice differentiable, one has:
\[
D_{f}(\lambda'|\lambda)=\frac{f''(1)\FI(\lambda)}{2}\times(\lambda'-\lambda)^{2}+\mathcal{O}\left((\lambda'-\lambda)^{3}\right),
\]
where $\FI(\lambda) = \mathrm{Var}_{\mu_\lambda}\left[s(X)\right]$ is the Fisher information.
\end{proposition}
This proposition applies in particular to the KL divergence ($f(x)=x\log x$, $f''(1)=1$), the reverse KL ($f(x)=-\log x$, $f''(1)=1$), all $\alpha-$divergences ($f''(1)=1$), the $\chi^{2}-$divergence ($f(x)=(x-1)^{2}$, $f''(1)=2$), hence fixing the $\chi^2$-divergence constant or the KL between consecutive iterates only differs by a multiplicative factor (resp. $\beta$ or $\nicefrac{\beta}{2})$.

The tempering strategy previously described can be justified by looking at the convergence of the corresponding entropic mirror descent scheme on the KL \eqref{eq:md_temp} where both $\cF$ and $B_{\phi}$ are chosen as the KL divergence.
Indeed, as shown in Eq.~\eqref{eq:kl_decreasing} in Appendix~\ref{sec:proof_th_rate}, the (Bregman) divergence between iterates $B_{\phi}(\mu_{n-1}|\mu_{n})=\KL(\mu_{n-1}|\mu_n)$ provides a lower bound on the decay of the objective $\cF(\mu)=\KL(\mu|\pi)$  achieved by one iteration of mirror descent : since $\gamma_n\leq L^{-1} = 1$ for all $n$, we have
\begin{align}\label{eq:kl_decrease}
    \KL(\mu_{n-1}|\pi) - \KL(\mu_n|\pi) 
    \geq \KL(\mu_{n-1}|\mu_n)  =\frac{\beta}{2}.
\end{align}

\Cref{prop:fdiv} above suggests the following recipe to choose successive
$\lambda_{n}$ values: 
\begin{equation}
\lambda_{n}-\lambda_{n-1}=c\FI(\lambda_{n-1})^{-1/2}\label{eq:lambda_sequence}
\end{equation}
for a certain $c=>0$; we recover $\chi^2(\mu_{n-1}|\mu_n)=\beta$ taking $c=\sqrt{\beta}$ 
(standard practice is to take $\beta=1$, which is equivalent to $\ess=N/2$). 
For a model where $\pi$
and $\mu_0$ correspond to the distribution of $d$ i.i.d. components, it is well known that $\FI(\lambda)=d\FI_{1}(\lambda)$ where $\FI_1$ denotes the Fisher information corresponding to one component.
We automatically get therefore that the number of successive
steps to move from $\lambda_0=0$ to $\lambda_T=1$ should be $\mathcal{O}(d^{1/2})$, as already observed by \citet[Proposition 17.2]{chopin2020introduction} and \citet[Section 3.3]{dai2022invitation} in the context of SMC (for Gaussian targets) and \citet[Section 2.3]{atchade2011towards} in the context of PT. 
See Figure~\ref{fig:tempering_seq} for a numerical experiment illustrating this point.


\subsection{Examples of tempering sequences}\label{sub:examples_sequences}

We now consider the simplified scenario in which $\pi$
and $\mu_0$ correspond to the distribution of $d$ i.i.d. components.
For some examples of proposals $\mu_0$ and targets $\pi$ the optimal tempering sequence satisfying~\eqref{eq:lambda_sequence} can be found (at least for large $d$) analytically. Our aim is to use the correspondence between $\gamma_n$ and $\lambda_n$ to obtain the convergence rate of the corresponding mirror descent scheme.

For large $d$, it makes sense to replace the sequence
$\lambda_{n}$ by a continuous function $\lambda(t)$, and solve the 
ODE:
\begin{equation}
\label{eq:ode}\dot{\lambda}=c\FI(\lambda)^{-1/2}.
\end{equation}
As a first simple case, consider targeting a pair of Gaussians with the same mean but different variances. Let $\pi = \mathcal{N}(0,\tau^{2}\Id)$ starting from $\mu_0 =\mathcal{N}(0,\Id) $. In this case $s(x)= x^{2}(1-1/\tau^{2})/2-\log \tau$, and we have $\mu_{\lambda}=\mathcal{N}\left(0,\left(1-\lambda+\lambda/\tau^{2}\right)^{-1}\Id\right)$ and $\FI(\lambda)\propto\left(1-\lambda+\lambda/\tau^{2}\right)^{-2}$. The corresponding ODE is  $
\dot{\lambda}=c(1+\alpha\lambda)$
with $\alpha=1/\tau^{2}-1$. If $\tau>1$, then $\alpha<0$, and the solution
is $\lambda(t)=1-\exp(\alpha t)$, which behaves likes a \emph{negative
exponential}. This corresponds to a constant $\gamma = 1-\exp(\alpha)$. 
Conversely, if $\tau<1$, then $\alpha>0,$ and the
solution is $\lambda(t)=\exp(\alpha t)-1$, which corresponds to \emph{exponential
growth}.

We then consider the case in which the variance is the same, but the means are different. Let $\mu_0=\mathcal{N}(0,\Id)$ and $\pi = \mathcal{N}(m,\Id)$, so $\mu_{\lambda}$ is $\mathcal{N}(\lambda m,\Id)$, $s(x)=mx-m^2/2$ , and $\FI(\lambda)=m^2$ is constant. In this case, $\lambda(t) = mt$ grows \emph{linearly}.

\begin{figure*}
    \centering
\begin{tikzpicture}[every node/.append style={font=\normalsize}]
\node (img1) {\includegraphics[width = 0.25\textwidth]{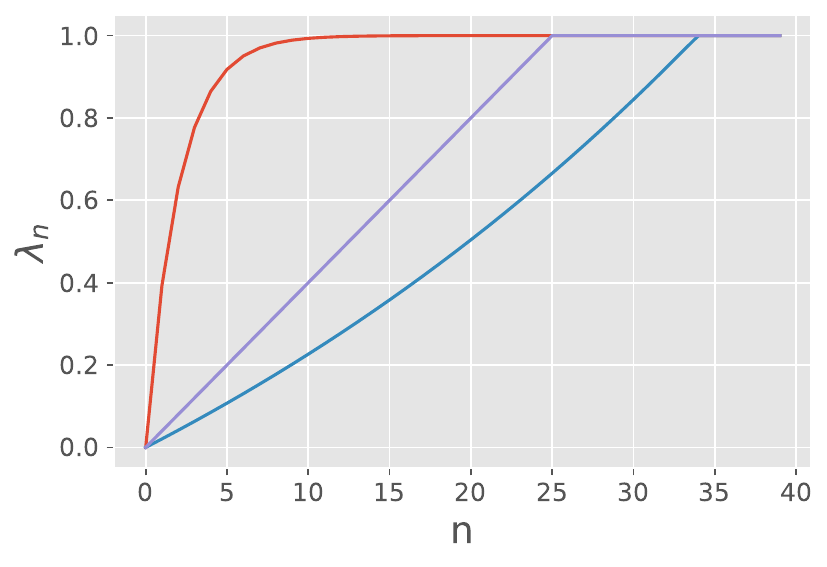}};
\node[right=of img1, node distance = 0, xshift = -0.7cm] (img2) {\includegraphics[width = 0.25\textwidth]{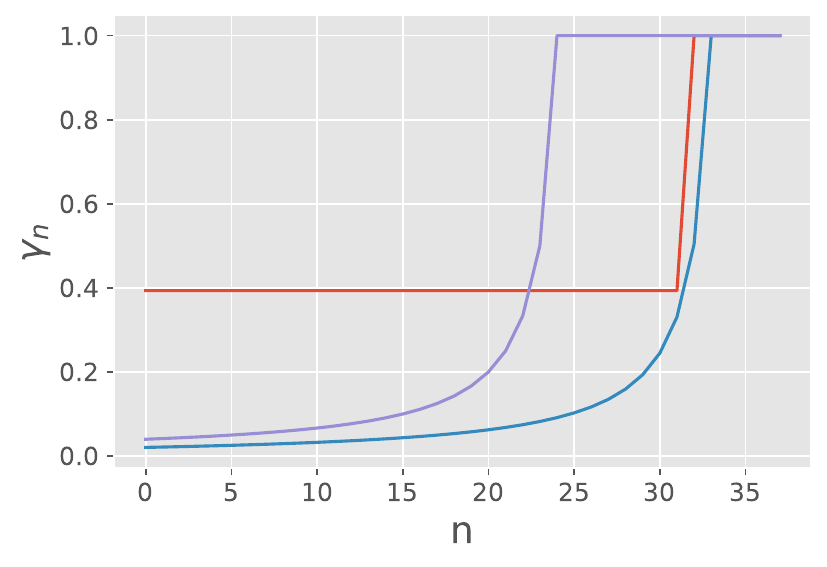}};
\node[right=of img2, node distance = 0, xshift = -0.7cm] (img3) {\includegraphics[width = 0.25\textwidth]{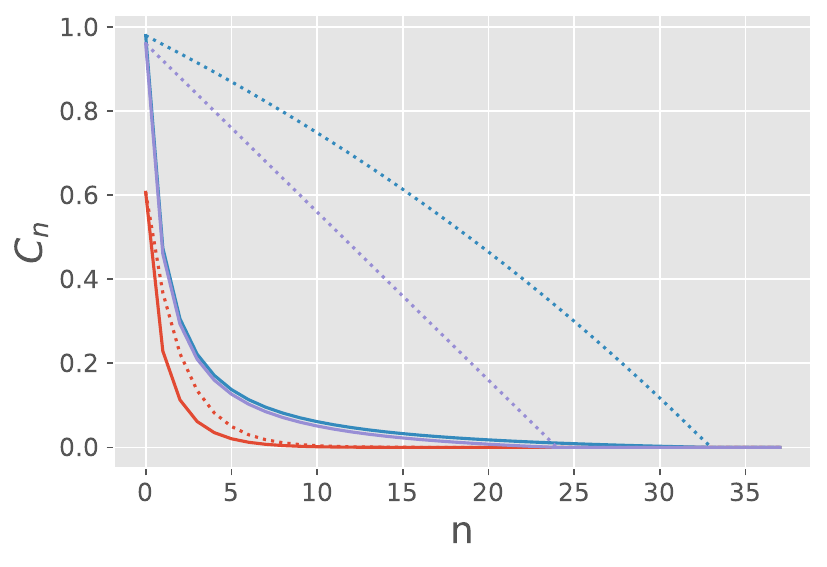}};
\node[below=of img1, node distance = 0, yshift = 1.2cm, xshift = 5cm] (img9) {\includegraphics[width = 0.25\textwidth]{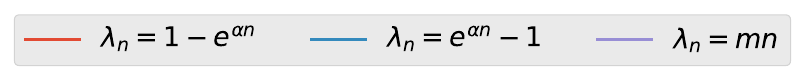}};
\end{tikzpicture}
    \caption{Sequence of $(\lambda_n)_{n\ge 1}$, $(\gamma_n)_{n\ge 1}$ and rate $C_n$ for the negative exponential, positive exponential and linear evolution of $(\lambda_n)_{n\ge 1}$. The dotted lines in the right-most plot show the bound $1-\lambda_n$ on $C_n$.}
    \label{fig:rates}
\end{figure*}

For each of the examples of tempering sequences $(\lambda_n)_{n\ge 1}$ exhibited in this subsection, we have seen at the beginning of this section that the tempered sequence converges at a rate $C_n \le 1- \lambda_n$.  Figure~\ref{fig:rates} provide some illustrations of the joint evolutions of temperatures $(\lambda_n)_{n\ge 1}$, mirror descent step sizes $(\gamma_n)_{n\ge 1}$ and rate of convergence $(C_n)_{n\ge 1}$ in these three different Gaussian scenarios. 

\section{Algorithmic approximations}\label{sec:algorithms}

Having identified the connection between the mirror descent iterates~\eqref{eq:md_temp} and the tempering iterates~\eqref{eq:tempering}, we now turn to existing (and potentially improvable) algorithms approximations, and identify their connections. 
Indeed, while \eqref{eq:md_temp} is attractive for its nice convergence properties, it is not feasible to run this iteration in practice for several reasons: each iteration depends on
the whole densities, and it requires a normalization step.

Notice from \eqref{eq:md} that it is natural to approximate entropic mirror descent on $\cF = \KL(\cdot|\pi)$ by
\begin{equation}\label{eq:approximate_scheme}
    \mu_{n+1} \propto q_n \exp(-\gamma_n g_n)
\end{equation} 
where $g_n$ is an approximation of the gradient of the KL objective $\log(\mu_n/\pi)$; and $q_n$ is an approximation of $\mu_n$.
We discuss here a common strategy in the Monte Carlo literature to approximate~\eqref{eq:tempering} based on importance sampling and show that the exponential update is performed on the importance weights.



Sequential Monte Carlo (SMC) samplers \citep{del2006sequential} provide a particle approximation of the tempering iterates~\eqref{eq:tempering} using clouds of $N$ weighted particles $\{X_n^i, W_n^i\}_{i=1}^N$. The fundamental ingredients of an SMC sampler are the sequence $(\lambda_n)_{n= 0}^T$  with $0= \lambda_0<\dots<\lambda_T =1$, a family of Markov kernels $(M_n)_{n= 1}^T$ used to propagate the particles forward in time and a resampling scheme.

For simplicity, we focus here on the case in which the Markov kernels $M_n$ are $\mu_n$-invariant, the resulting SMC algorithm is summarized in Algorithm~\ref{alg:smc} in \Cref{sec:algorithm_details}.
At iteration $n$ the weighted particle set $\{X_{n-1}^i, W_{n-1}^i\}_{i=1}^N$ is resampled to obtain the equally weighted particle set $\{\widetilde{X}_{n-1}^i, 1/N\}_{i=1}^N$ and the kernel $M_{n}$ is applied to propose new particle locations $X_{n}^i\sim M_{n}(\cdot, \widetilde{X}_{n-1}^i)$. The weights are proportional to
\begin{align}\label{eq:smc_weight}
w_{n}(x)=\frac{\eta_{n}(x)}{\eta_{n-1}(x)} =\left(\frac{\pi(x)}{\mu_0(x)}\right)^{\lambda_{n}-\lambda_{n-1}}
\end{align}
where $\eta_{n}:=\mu_0^{1-\lambda_{n}}\pi^{\lambda_{n}}$ and $\mu_{n} = \eta_{n}/Z_{n}$. Recalling the relationship between the sequence of $\gamma_n$ and of $\lambda_n$, $\gamma_n = (\lambda_n - \lambda_{n-1})/(1-\lambda_{n-1})$, we find that
\begin{align}
\label{eq:smc_pmd_weight}
    w_{n}(x)
    &= \left(\frac{\pi(x)}{\eta_{n-1}(x)}\right)^{\gamma_{n}} \propto \left(\frac{\pi(x)}{\mu_{n-1}(x)}\right)^{\gamma_{n}},
\end{align}
where the normalizing constant can be discarded due to the re-normalization,
showing that the importance weights used within an SMC sampler approximate the exponential update  in~\eqref{eq:approximate_scheme}.
The approximation of $
\mu_{n}$ provided by SMC is $q_n^{\textrm{SMC}}(x) =\sum_{i=1}^N W_n^i\delta_{X_n^i}(x)$, where $\delta(\cdot)$ denotes the Dirac's delta function and $W_n^i = w_n(\widetilde{X}_{n-1}^i)$. 

\begin{remark}
In \Cref{sec:algorithm_details} we discuss two alternative strategies to SMC based on importance sampling that directly approximate \eqref{eq:md} \citep{dai2016provable, korba2022adaptive}. 
We highlight in particular that MD on measures can be implemented through SMC with a better complexity than the scheme proposed in \cite{dai2016provable} (the weights~\eqref{eq:smc_weight} do not depend on the $N$ particle set and can be computed in $\mathcal{O}(1)$ time while those of \cite{dai2016provable} depend on the $N$ particle set and require $\mathcal{O}(N)$ cost, see \Cref{sec:algorithm_details}).
\end{remark}

In the SMC literature, the tempering schedule $\{\lambda_n\}_{n=1}^T$ is normally chosen adaptively, by ensuring that the $\chi^2$ divergence between successive distributions is constant and sufficiently small.
The $\chi^2$ divergence is approximated as $\chi^2(\mu_{n-1}|\mu_n)\approx  \frac{N}{\ess_n(\lambda_n)}-1$ (see, e.g. \citet[Section 8.6]{chopin2020introduction} for a justification),
where $\ess_n$ denotes the effective sample size 
\begin{align*}
    \ess_n(\lambda)  &:=  \left(\sum_{i=1}^N w_n(\widetilde{X}_{n-1}^i)\right)^2 /\sum_{i=1}^N (w_n(\widetilde{X}_{n-1}^i))^2.
\end{align*}
Given $\beta>0$, in standard adaptive SMC we need to solve $\ess_n(\lambda) = N/(\beta+1)$ at each iteration $n$. This is normally achieved via the bisection method, since $\ess_n(\lambda)$ is a nonlinear function of $\lambda$ taking values in $[1, N]$. 

We conducted a numerical experiment to study the possible behaviors of the tempering sequences found by such adaptive strategies when using SMC samplers. Our numerical results are obtained using waste-free SMC \citep{dau2022waste}, as there is evidence that it improves on standard SMC, which in turns outperforms annealed importance sampling \citep{jasra2011inference}. For this experiment we use Markov kernels that are random-walk Metropolis kernels, automatically calibrated on the current particle sample; the next tempering exponent is chosen so that $\ess_n=N/2$, and $N=10^4$ and $d=25$. The code will be made publicly available.

\begin{figure}
\begin{centering}
\includegraphics[scale=0.25]{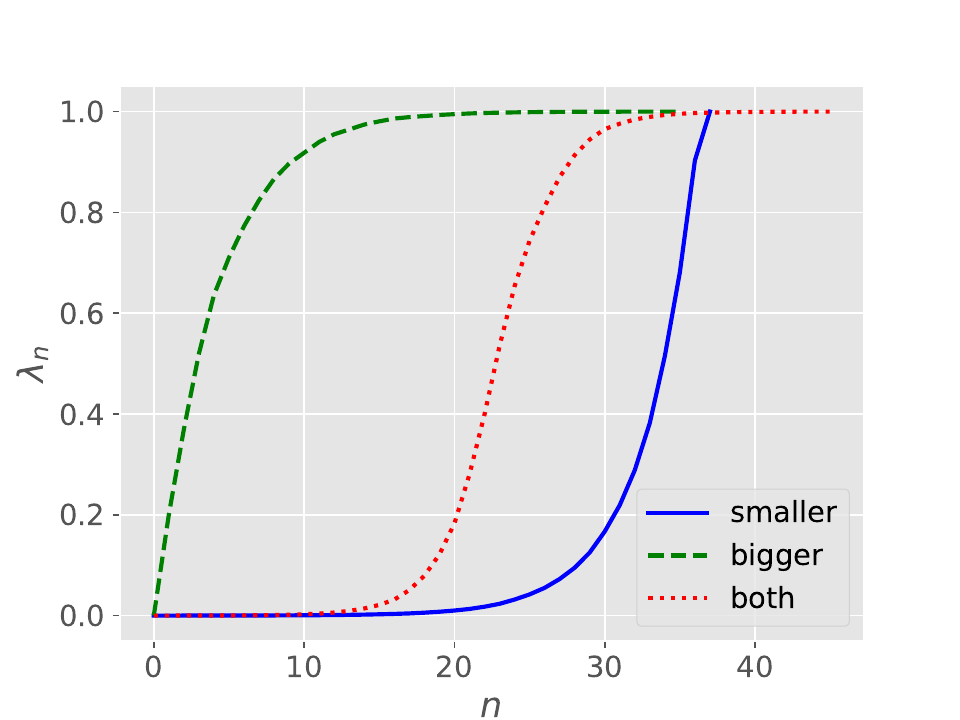}
\includegraphics[scale=0.25]{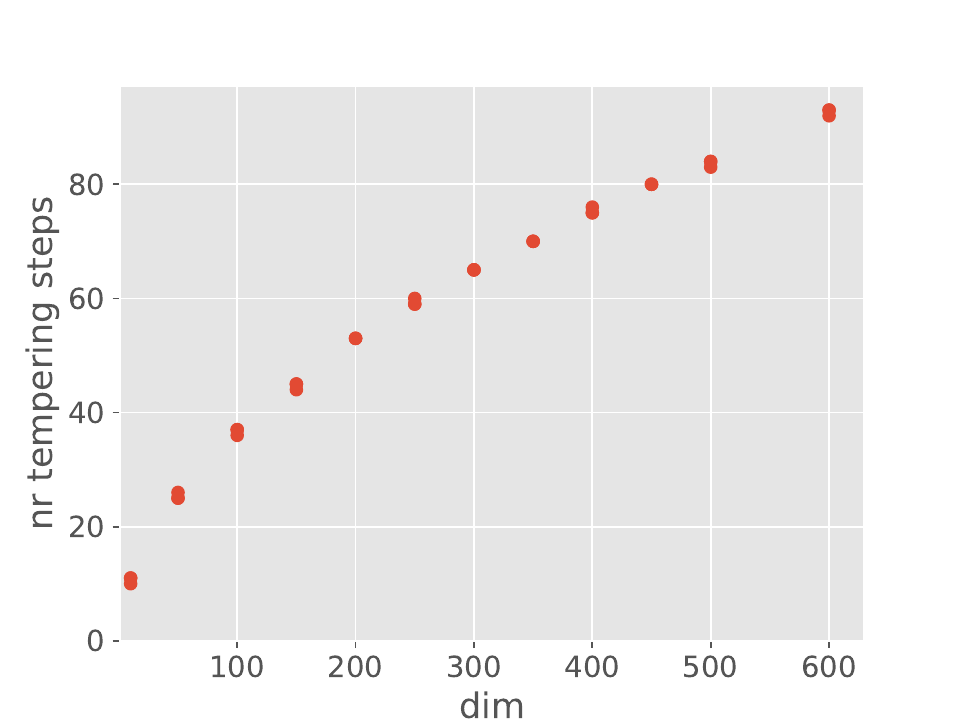}
\par\end{centering}
\caption{\label{fig:tempering_seq}Left: adapted tempering sequences for different $\Sigma$.
Right: Length
of tempering sequence as a function of $d$ in scenario (a); one recovers
the $\mathcal{O}(d^{1/2})$ scaling. }
\end{figure}

\Cref{fig:tempering_seq} (left) plots the tempering sequence $(\lambda_{n})_{n\ge 1}$ computed adaptively on a toy example where $\mu_{0}=\mathcal{N}_{d}(0_{d},\Id)$,
$\pi=\mathcal{N}_{d}(m,\Sigma)$, $m=1_{d}$, and various choices for $\Sigma$: 
(a) $\Sigma=10^{-2}\Id$; (b) $\Sigma=10^{2}\Id$; and (c) $\Sigma=\mathrm{diag}(v)$,
with the first $(d/2)$ elements of $v$ equal to $10^{-2},$ and
the remaining elements equal to $10^{2}$.  Cases (a) and (b) illustrates our theoretical tempering rates of  Section~\ref{sub:examples_sequences}
; when the target has smaller (resp. larger) variance along all directions, the tempering
sequence behaves like a positive (resp. negative) exponential. Case (c) is particularly interesting
as it shows that the tempering sequence may behave as a mix between the two cases; when the variance of the target is both larger in certain directions and smaller in other directions (relative to $\mu_0$), then 
the tempering sequence must slow down both at the beginning and at the end. The bottom line of this experiment is that what constitutes a "good" tempering sequence varies strongly according to the pair $(\mu_0, \pi)$, and thus using an adaptive strategy is essential for good
performance.

\Cref{fig:tempering_seq} (right) plots the number of tempering steps obtained from the algorithm as a function of $d$, in the "smaller variance" scenario, $\Sigma=10^{-2}\Id$. One recovers the $\mathcal{O}(d^{1/2})$ scaling derived in Section~\ref{sec:strategy_tempered}. The reader may refer to \Cref{sec:algorithm_details} for more details on the implementation. 

\begin{remark}
Proposition~\ref{prop:fdiv} provides alternative methods to approximately solve $\chi^2(\mu_{n-1}|\mu_n)=\beta$ in a equivalent way (up to higher order terms) from the current set of particles. A first one is to fix to a constant:
\begin{align*}
    \KL(\mu_{n-1}|\mu_n) 
    \approx -\frac{1}{N}\sum_{i=1}^N\log w_n(\widetilde{X}_{n-1}^i)
    + \log \frac{1}{N}\sum_{i=1}^Nw_n(\widetilde{X}_{n-1}^i),
\end{align*}
which is likely to be more stable than the $\ess$, since it involves the log-weights rather than the weights themselves. The second one is set the next $\lambda_n$ as $\lambda_n = \lambda_{n-1} + (\beta /\hat{\FI}(\lambda_{n-1}))^{1/2}$, where $\hat{\FI}(\lambda_{n-1}) $ is
\begin{align*}
      \frac{1}{N}\sum_{i=1}^N \left(\log \frac{\pi(\widetilde{X}_{n-1}^i)}{\mu_0(\widetilde{X}_{n-1}^i)}\right)^2
  - \left(\frac{1}{N}\sum_{i=1}^N \log \frac{\pi(\widetilde{X}_{n-1}^i)}{\mu_0(\widetilde{X}_{n-1}^i)}\right)^2.
\end{align*}
\end{remark}


\section{Related work}\label{sec:relevant_work}

In this section we discuss alternative tempering strategies and algorithmic approximations related to the tempering update.

\textbf{Tempering, KL divergence optimization and normalizing constant estimation.}

The insight given by the mirror descent perspective allows us to relate sampling and estimation of the normalizing constant $\mathcal{Z}$ of $\pi$. In the AIS literature, the optimal sequence of distributions $(\mu_n)_{n\geq 1}$ is normally chosen to minimize the bias of the log-weights \cite{grosse2013annealing, goshtasbpour2023adaptive}
\begin{align*}
    \log \mathcal{Z} - \mathbb{E}[\log w_n] = \sum_{n=1}^T \KL(\mu_{n-1}|\mu_n).
\end{align*}
Assuming the first $(n-1)$ iterates are fixed, one finds $\mu_n$ by minimizing $\KL(\mu_{n-1}|\mu_n)$, which corresponds to the approach adopted in the SMC literature described above.

\cite{grosse2013annealing} derive optimal paths for exponential families and show that
\begin{align*}
    \mu_\lambda = \arg\min_{\mu} \left[(1-\lambda)\KL(\mu|\mu_0)+\lambda\KL(\mu|\pi)\right].
\end{align*}
This corresponds to the first step of entropic mirror descent, i.e. \eqref{eq:prox_algo} with $n=0$ and $\lambda = \gamma_1$.

\begin{figure}
\begin{centering}
\includegraphics[scale=0.25]{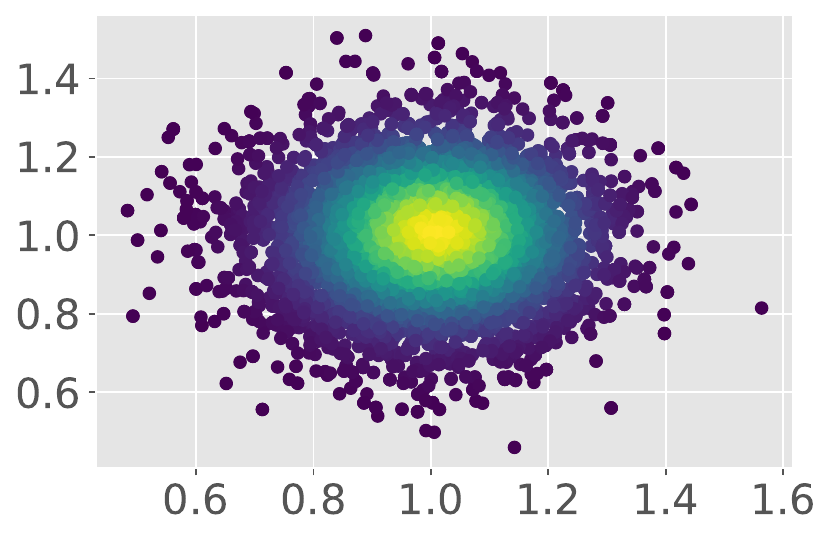}
\includegraphics[scale=0.25]{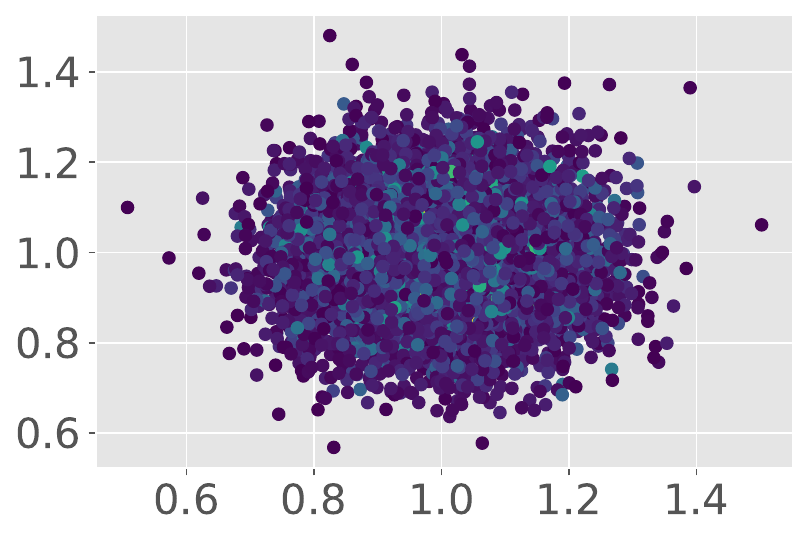}
\par\end{centering}
\caption{\label{fig:smc_vs_ais}Approximations of $\pi = \mathcal{N}(1_d, 0.1^2\Id)$. Left: SMC, 5 tempering steps.
Right: \cite{goshtasbpour2023adaptive}, 6257 tempering steps.  }
\end{figure}

\citet[Proposition 3.2]{goshtasbpour2023adaptive} show that the tempering sequence is the path of steepest descent for the KL; i.e. that minimizes \eqref{eq:first_var} infinitesimally, where the perturbation $\xi$ is a smooth ($C^1$) perturbation with bounded variance. Given $(n-1)$ tempering iterates, they select the next one minimizing $\KL(\mu_n|\pi)$ instead and identify a tempering schedule that decreases this objective with constant rate and satisfies the ODE
\begin{equation}
\label{eq:ode_other_paper}
 \dot{\lambda} =    c\left[\FI(\lambda)(1-\lambda)\right]^{-1}.
\end{equation}
This differs from ours in \eqref{eq:ode} which keeps the $\KL$ between successive entropic mirror descent iterates constant. Both strategies can be justified using the well known identity (\citet[Section 4.4]{brekelmans2020all} and \Cref{app:odes})
\begin{align}\label{eq:kl_integration}
\KL(\mu_{\lambda}|\mu_{\lambda'}) &= 
    \int_{\lambda}^{\lambda'}(\lambda-u)\FI(u) du.\nonumber
\end{align}
Using~\eqref{eq:kl_decrease}, we find that
\begin{align*}
0 \leq \KL(\mu_{n-1}|\mu_n) \leq \KL(\mu_{n-1}|\pi) - \KL(\mu_n|\pi),
\end{align*}
which shows that the \eqref{eq:ode} and \eqref{eq:ode_other_paper}  fulfil opposite goals, i.e. \eqref{eq:ode_other_paper} aims to find $\mu_n$ which minimizes $\KL(\mu_n|\pi)$ i.e. an upper bound on $\KL(\mu_{n-1}|\mu_n)$.
From the numerical point of view, both strategies employ the importance weights to select the next $\lambda$; however, in our strategy the weights are those obtained by importance sampling with target $\mu_{\lambda_n}$ and proposal $\mu_{\lambda_{n-1}}$, in \cite{goshtasbpour2023adaptive} the target is $\pi$. As a consequence, their method is more numerically unstable due to the higher variance of the weights. To see this we reproduce their narrow Gaussian experiment in Figure~\ref{fig:smc_vs_ais} and compare with waste-free SMC with the same setup of \Cref{sec:algorithms} (see \Cref{app:numerical_details} for implementation details). The target is $\pi = \mathcal{N}(1_d, 0.1^2\Id)$ and $\mu_0 = \mathcal{N}(0_d, \Id)$. SMC better approximates $\pi$ and requires only 5 tempering steps, while \cite{goshtasbpour2023adaptive} provides worse approximations and require more than 6000 steps.

\cite{kiwaki2015variational} consider the variance of $\log w_n$ instead and derive an ODE similar to~\eqref{eq:ode}. Their method however requires running AIS twice to select the successive $\lambda$, while our rule~\eqref{eq:lambda_sequence} only requires to evaluate $\hat{\FI}(\lambda)$

An ODE involving $\FI(\lambda)$ was also derived in \cite{gelman1998simulating}, but, as mentioned by the authors, it often results in intractable optimal paths.

\textbf{Effect of the dimension on tempering/SMC in \cite{beskos2014stability}.}
In this paper the authors investigate the effect of dimension on the stability of SMC methods. There, stability refers to the ability of the SMC algorithm to produce accurate approximations of the target distribution as the dimension increases, while keeping the computational cost reasonable. The authors show that for a certain class of target densities (an i.i.d. target of the form $\pi(x) = \prod_{i=1}^d \pi^i(x^i)$ where $x=(x^1,\dots, x^d)\in \R^d$),  SMC  with the tempering sequence defined as $\lambda_n = \lambda_1 + (n-1)(1-\lambda_1)d^{-1}$, $1\le n\le d$, 
is stable, i.e. the $\ess$ converges weakly to a non-trivial limit ($\ess \in (1, N)$) as $d$ grows and the number of particles is kept fixed.
This result suggests using $\mathcal{O}(d)$ tempering iterations contrary to the $\mathcal{O}(d^{1/2})$ found in \Cref{prop:fdiv} and confirmed by our numerical results. We leave further investigation of the optimal scaling with $d$ of SMC samplers for future work.

\textbf{Parallel tempering.} The tempering iterates~\eqref{eq:tempering} are also at the basis of Parallel Tempering (PT) \citep{geyer1991markov,hukushima1996exchange}, a class of Markov chain Monte Carlo algorithms which relies on interacting Markov chains to sample from~\eqref{eq:tempering}. 
PT is not based on importance sampling, hence the connection with Mirror Descent is less clear since identifying an update of the form~\eqref{eq:approximate_scheme} is not possible. It is customary in PT \citep[e.g. Section 4]{syed2022non} to fix the tempering sequences so that the acceptance probability of a swap between two successive $\lambda_n$ is constant in $n$. 
One may use Proposition 1 
of \citet{predescu2004incomplete}, see also Theorem 2 in \citet{syed2022non}, which is similar in spirit to 
our Proposition~\ref{prop:fdiv}, but not equivalent: Proposition~\ref{prop:fdiv} applies to the $f$ divergence between $\mu_{\lambda}$
and $\mu_{\lambda'}$, for $\lambda'\approx\lambda$, where $f$ is differentiable, whereas the acceptance rate 
of a PT swap is the TV distance between $\mu_\lambda\otimes\mu_{\lambda'}$ and $\mu_{\lambda'}\otimes\mu_{\lambda}$, again for  $\lambda'\approx\lambda$.
Moreover, the TV distance is a $f$-divergence with $f(t)=|t-1|$, which is not differentiable at 1. 

\textbf{Adaptive tempering.}
In \cite{korba2022adaptive}, the authors propose to choose the step-size $\gamma_n$ as follows. At time $n$ draw $m_n$ particles from $q_n^{\textrm{SRAIS}}$.
Let
$P = \sum_{l=1}^{m_n}u_{n}^{l}\delta_{X_{n}^{l}}$
 and 
 $Q= \sum_{l=1}^{m_n} (m_n)^{-1} \delta_{X_{n}^{l}}$ the reweighted and uniform distribution on the particles $(X_{n}^{l})_{l=1}^{m_n}$ respectively, where $u_{n}^{l} = u_n(X_{n}^l) = \pi(x)/q_n(x)$ are the importance weights between the target distribution $\pi$ and the current approximate iterate $q_n$ of $\mu_n$. 
\cite{korba2022adaptive} propose to set 
$\gamma_n$ as $\gamma_{n} = 1 - R_\alpha (P | Q)/\log(m_n)$, 
where $R_\alpha$ is Renyi's $\alpha$-divergence \citep{renyi1961measures} of $P$ from $Q$, 
in particular $R_1 (P || Q)= \KL(P|Q)$, and $\log(m_n)$ normalizes the ratio between 0 and 1. Hence, for $\alpha =1$ and without the discrete particle approximation, their rule can be written $\gamma_n = 1- \KL(\mu_n|\pi).$ Since $\gamma_n = (\lambda_n-\lambda_{n-1})/(1-\lambda_{n-1})$, by the decrease of KL formula \eqref{eq:kl_decrease}, this rule can also be seen as enforcing a gap between consecutive $\lambda$'s as a constant.



\section{Conclusions}

This paper establishes a connection between entropic mirror descent and tempering to sample from a target probability distribution known up to a normalizing constant. 
We show that the two strategies are equivalent 
and obtain an explicit convergence rate 
for the tempering iterates.  This convergence rate does not depend on the convexity properties of the target $\pi$, contrary to the rates for Langevin Monte Carlo \cite{durmus2019analysis, vempala2019rapid, karimi2016linear}. 

We provide an optimization point of view on sequential Monte Carlo by identifying the SMC update as \eqref{eq:approximate_scheme} and motivate the adaptive strategy commonly used in the literature through mirror descent. Furthermore, we identify that for a number of algorithms based on importance sampling, the importance weights carry the gradient information, and propose strategies to reduce their numerical error (see \Cref{sec:algorithm_details}). 
By comparing several approximations of entropic mirror descent and several adaptive strategies to select the sequence, we find that SMC has generally lower cost and the tempering rule \eqref{eq:ode} is more stable than alternatives.
This connection enabled us to tackle the selection of the tempering schedule in a principled way, and derive several conclusions that were not known in or contradicts the previous tempering literature, for instance that the length of the tempering schedule should scale as $\sqrt{d}$. Our approach yields a simpler and more numerically stable tempering rule than other schemes (minus the standard $\ess-$based rule in the SMC literature, which gives essentially the same results as ours). 

\section*{Impact statement}

Our paper has a theoretical and practical interest for the literature on sampling and MCMC algorithms. On the theoretical side, our study yields a rate of convergence for the target sequence that SMC algorithms are tracking. On the practical side, we show that common practices such as the ESS-based rules, are more theoretically grounded, simpler and more efficient than alternative proposals. This may have a substantial impact in the deployment of Bayesian inference tasks which rely on MCMC algorithms, and enable to predict with uncertainty.

\bibliographystyle{apalike}
\bibliography{biblio}

\begin{thebibliography}{}

\bibitem[Amari, 2016]{amari2016information}
Amari, S.-i. (2016).
\newblock {\em Information Geometry and Its Applications}, volume 194.
\newblock Springer.

\bibitem[Ambrosio et~al., 2008]{ambrosio2008gradient}
Ambrosio, L., Gigli, N., and Savar{\'e}, G. (2008).
\newblock {\em Gradient flows: in metric spaces and in the space of probability
  measures}.
\newblock Springer Science \& Business Media.

\bibitem[Atchad{\'e} et~al., 2011]{atchade2011towards}
Atchad{\'e}, Y.~F., Roberts, G.~O., and Rosenthal, J.~S. (2011).
\newblock {Towards optimal scaling of Metropolis-coupled Markov chain Monte
  Carlo}.
\newblock {\em Statistics and Computing}, 21:555--568.

\bibitem[Aubin-Frankowski et~al., 2022]{aubin2022mirror}
Aubin-Frankowski, P.-C., Korba, A., and L{\'e}ger, F. (2022).
\newblock {Mirror descent with relative smoothness in measure spaces, with
  application to Sinkhorn and EM}.
\newblock {\em Advances in Neural Information Processing Systems},
  35:17263--17275.

\bibitem[Beck and Teboulle, 2003]{beck2003mirror}
Beck, A. and Teboulle, M. (2003).
\newblock Mirror descent and nonlinear projected subgradient methods for convex
  optimization.
\newblock {\em Operations Research Letters}, 31(3):167--175.

\bibitem[Beskos et~al., 2014]{beskos2014stability}
Beskos, A., Crisan, D., and Jasra, A. (2014).
\newblock On the stability of sequential {Monte Carlo} methods in high
  dimensions.
\newblock {\em Annals of Applied Probability}, 24(4):1396--1445.

\bibitem[Brekelmans et~al., 2020]{brekelmans2020all}
Brekelmans, R., Masrani, V., Wood, F., Steeg, G.~V., and Galstyan, A. (2020).
\newblock All in the exponential family: {B}regman duality in thermodynamic
  variational inference.
\newblock In III, H.~D. and Singh, A., editors, {\em Proceedings of the 37th
  International Conference on Machine Learning}, volume 119 of {\em Proceedings
  of Machine Learning Research}, pages 1111--1122. PMLR.

\bibitem[Chac{\'o}n and Duong, 2018]{chacon2018multivariate}
Chac{\'o}n, J.~E. and Duong, T. (2018).
\newblock {\em Multivariate kernel smoothing and its applications}.
\newblock Chapman and Hall/CRC.

\bibitem[Chizat, 2022]{chizat2021convergence}
Chizat, L. (2022).
\newblock Convergence {Rates} of {Gradient} {Methods} for {Convex}
  {Optimization} in the {Space} of {Measures}.
\newblock {\em Open Journal of Mathematical Optimization}, 3.

\bibitem[Chopin and Papaspiliopoulos, 2020]{chopin2020introduction}
Chopin, N. and Papaspiliopoulos, O. (2020).
\newblock {\em An introduction to sequential Monte Carlo}.
\newblock Springer.

\bibitem[Crisan and Doucet, 2002]{smc:theory:CD02}
Crisan, D. and Doucet, A. (2002).
\newblock A survey of convergence results on particle filtering methods for
  practitioners.
\newblock {\em IEEE Transactions on Signal Processing}, 50(3):736--746.

\bibitem[Dai et~al., 2016]{dai2016provable}
Dai, B., He, N., Dai, H., and Song, L. (2016).
\newblock {Provable Bayesian inference via particle mirror descent}.
\newblock In {\em Artificial Intelligence and Statistics}, pages 985--994.
  PMLR.

\bibitem[Dai et~al., 2022]{dai2022invitation}
Dai, C., Heng, J., Jacob, P.~E., and Whiteley, N. (2022).
\newblock An invitation to sequential {Monte Carlo }samplers.
\newblock {\em Journal of the American Statistical Association},
  117(539):1587--1600.

\bibitem[Dau and Chopin, 2022]{dau2022waste}
Dau, H.-D. and Chopin, N. (2022).
\newblock {Waste-free sequential Monte Carlo}.
\newblock {\em Journal of the Royal Statistical Society Series B: Statistical
  Methodology}, 84(1):114--148.

\bibitem[Del~Moral, 2004]{smc:theory:Del04}
Del~Moral, P. (2004).
\newblock {\em {Feynman-Kac} formulae: genealogical and interacting particle
  systems with applications}.
\newblock Probability and Its Applications. Springer Verlag, New York.

\bibitem[Del~Moral et~al., 2006]{del2006sequential}
Del~Moral, P., Doucet, A., and Jasra, A. (2006).
\newblock {Sequential Monte Carlo samplers}.
\newblock {\em Journal of the Royal Statistical Society Series B: Statistical
  Methodology}, 68(3):411--436.

\bibitem[Domingo-Enrich and Pooladian, 2023]{domingo2023explicit}
Domingo-Enrich, C. and Pooladian, A.-A. (2023).
\newblock {An Explicit Expansion of the Kullback-Leibler Divergence along its
  Fisher-Rao Gradient Flow}.
\newblock {\em arXiv preprint arXiv:2302.12229}.

\bibitem[Durmus et~al., 2019]{durmus2019analysis}
Durmus, A., Majewski, S., and Miasojedow, B. (2019).
\newblock {Analysis of Langevin Monte Carlo via convex optimization}.
\newblock {\em The Journal of Machine Learning Research}, 20(1):2666--2711.

\bibitem[Garbuno-Inigo et~al., 2020]{garbuno2020interacting}
Garbuno-Inigo, A., Hoffmann, F., Li, W., and Stuart, A.~M. (2020).
\newblock {Interacting Langevin diffusions: Gradient structure and ensemble
  Kalman sampler}.
\newblock {\em SIAM Journal on Applied Dynamical Systems}, 19(1):412--441.

\bibitem[Gelman and Meng, 1998]{gelman1998simulating}
Gelman, A. and Meng, X.-L. (1998).
\newblock Simulating normalizing constants: From importance sampling to bridge
  sampling to path sampling.
\newblock {\em Statistical Science}, pages 163--185.

\bibitem[Geyer, 1991]{geyer1991markov}
Geyer, C.~J. (1991).
\newblock {Markov chain Monte Carlo maximum likelihood}.
\newblock In Keramides, E.~M., editor, {\em Computing Science and Statistics:
  Proceedings of the 23rd Symposium on the Interface}, pages 156--163.

\bibitem[Goshtasbpour et~al., 2023]{goshtasbpour2023adaptive}
Goshtasbpour, S., Cohen, V., and Perez-Cruz, F. (2023).
\newblock Adaptive annealed importance sampling with constant rate progress.
\newblock In Krause, A., Brunskill, E., Cho, K., Engelhardt, B., Sabato, S.,
  and Scarlett, J., editors, {\em Proceedings of the 40th International
  Conference on Machine Learning}, volume 202 of {\em Proceedings of Machine
  Learning Research}, pages 11642--11658. PMLR.

\bibitem[Grosse et~al., 2013]{grosse2013annealing}
Grosse, R.~B., Maddison, C.~J., and Salakhutdinov, R.~R. (2013).
\newblock Annealing between distributions by averaging moments.
\newblock In Burges, C., Bottou, L., Welling, M., Ghahramani, Z., and
  Weinberger, K., editors, {\em Advances in Neural Information Processing
  Systems}, volume~26. Curran Associates, Inc.

\bibitem[Hukushima and Nemoto, 1996]{hukushima1996exchange}
Hukushima, K. and Nemoto, K. (1996).
\newblock {Exchange Monte Carlo method and application to spin glass
  simulations}.
\newblock {\em Journal of the Physical Society of Japan}, 65(6):1604--1608.

\bibitem[Jasra et~al., 2011]{jasra2011inference}
Jasra, A., Stephens, D.~A., Doucet, A., and Tsagaris, T. (2011).
\newblock {Inference for L{\'e}vy-driven stochastic volatility models via
  adaptive sequential Monte Carlo}.
\newblock {\em Scandinavian Journal of Statistics}, 38(1):1--22.

\bibitem[Jordan et~al., 1998]{jordan1998variational}
Jordan, R., Kinderlehrer, D., and Otto, F. (1998).
\newblock The variational formulation of the {Fokker--Planck} equation.
\newblock {\em SIAM Journal on Mathematical Analysis}, 29(1):1--17.

\bibitem[Karimi et~al., 2016]{karimi2016linear}
Karimi, H., Nutini, J., and Schmidt, M. (2016).
\newblock {Linear convergence of gradient and proximal-gradient methods under
  the Polyak-{\L}ojasiewicz condition}.
\newblock In {\em Machine Learning and Knowledge Discovery in Databases:
  European Conference, ECML PKDD 2016, Riva del Garda, Italy, September 19-23,
  2016, Proceedings, Part I 16}, pages 795--811. Springer.

\bibitem[Kiwaki, 2015]{kiwaki2015variational}
Kiwaki, T. (2015).
\newblock Variational optimization of annealing schedules.
\newblock {\em arXiv preprint arXiv:1502.05313}.

\bibitem[Korba and Portier, 2022]{korba2022adaptive}
Korba, A. and Portier, F. (2022).
\newblock Adaptive importance sampling meets mirror descent: a bias-variance
  tradeoff.
\newblock In {\em International Conference on Artificial Intelligence and
  Statistics}, pages 11503--11527. PMLR.

\bibitem[Liu, 2017]{liu2017stein}
Liu, Q. (2017).
\newblock Stein variational gradient descent as gradient flow.
\newblock In Guyon, I., Luxburg, U.~V., Bengio, S., Wallach, H., Fergus, R.,
  Vishwanathan, S., and Garnett, R., editors, {\em Advances in Neural
  Information Processing Systems}, volume~30. Curran Associates, Inc.

\bibitem[Lu et~al., 2018]{lu2018relatively}
Lu, H., Freund, R.~M., and Nesterov, Y. (2018).
\newblock Relatively smooth convex optimization by first-order methods, and
  applications.
\newblock {\em SIAM Journal on Optimization}, 28(1):333--354.

\bibitem[Lu et~al., 2023]{lu2023birth}
Lu, Y., Slep{\v{c}}ev, D., and Wang, L. (2023).
\newblock Birth--death dynamics for sampling: global convergence,
  approximations and their asymptotics.
\newblock {\em Nonlinearity}, 36(11):5731.

\bibitem[Mou et~al., 2021]{mou2021high}
Mou, W., Ma, Y.-A., Wainwright, M.~J., Bartlett, P.~L., and Jordan, M.~I.
  (2021).
\newblock {High-order Langevin diffusion yields an accelerated MCMC algorithm}.
\newblock {\em The Journal of Machine Learning Research}, 22(1):1919--1959.

\bibitem[Neal, 2001]{neal2001annealed}
Neal, R.~M. (2001).
\newblock Annealed importance sampling.
\newblock {\em Statistics and Computing}, 11:125--139.

\bibitem[Otto and Villani, 2000]{otto2000generalization}
Otto, F. and Villani, C. (2000).
\newblock Generalization of an inequality by {Talagrand} and links with the
  logarithmic {Sobolev} inequality.
\newblock {\em Journal of Functional Analysis}, 173(2):361--400.

\bibitem[Predescu et~al., 2004]{predescu2004incomplete}
Predescu, C., Predescu, M., and Ciobanu, C.~V. (2004).
\newblock The incomplete beta function law for parallel tempering sampling of
  classical canonical systems.
\newblock {\em The Journal of Chemical Physics}, 120(9):4119--4128.

\bibitem[R{\'e}nyi et~al., 1961]{renyi1961measures}
R{\'e}nyi, A. et~al. (1961).
\newblock On measures of entropy and information.
\newblock In {\em Proceedings of the Fourth Berkeley Symposium on Mathematical
  Statistics and Probability, Volume 1: Contributions to the Theory of
  Statistics}. The Regents of the University of California.

\bibitem[Roberts and Tweedie, 1996]{roberts1996exponential}
Roberts, G.~O. and Tweedie, R.~L. (1996).
\newblock {Exponential convergence of Langevin distributions and their discrete
  approximations}.
\newblock {\em Bernoulli}, pages 341--363.

\bibitem[Salim et~al., 2020]{salim2020wasserstein}
Salim, A., Korba, A., and Luise, G. (2020).
\newblock {The Wasserstein proximal gradient algorithm}.
\newblock {\em Advances in Neural Information Processing Systems},
  33:12356--12366.

\bibitem[Silverman, 1986]{silverman1986density}
Silverman, B.~W. (1986).
\newblock {\em Density Estimation for Statistics and Data Analysis}.
\newblock Chapman \& Hall.

\bibitem[Syed et~al., 2022]{syed2022non}
Syed, S., Bouchard-C{\^o}t{\'e}, A., Deligiannidis, G., and Doucet, A. (2022).
\newblock {Non-reversible parallel tempering: a scalable highly parallel MCMC
  scheme}.
\newblock {\em Journal of the Royal Statistical Society Series B: Statistical
  Methodology}, 84(2):321--350.

\bibitem[Syed et~al., 2021]{syed2021parallel}
Syed, S., Romaniello, V., Campbell, T., and Bouchard-C{\^o}t{\'e}, A. (2021).
\newblock Parallel tempering on optimized paths.
\newblock In {\em International Conference on Machine Learning}, pages
  10033--10042. PMLR.

\bibitem[Vempala and Wibisono, 2019]{vempala2019rapid}
Vempala, S. and Wibisono, A. (2019).
\newblock {Rapid convergence of the unadjusted Langevin algorithm: Isoperimetry
  suffices}.
\newblock {\em Advances in neural information processing systems}, 32.

\bibitem[Wibisono, 2018]{wibisono2018sampling}
Wibisono, A. (2018).
\newblock {Sampling as optimization in the space of measures: The Langevin
  dynamics as a composite optimization problem}.
\newblock In {\em Conference on Learning Theory}, pages 2093--3027. PMLR.

\bibitem[Ying, 2020]{ying2020mirror}
Ying, L. (2020).
\newblock Mirror descent algorithms for minimizing interacting free energy.
\newblock {\em Journal of Scientific Computing}, 84(3):51.

\end{thebibliography}
\newpage
\appendix 
\onecolumn
\section{Proof of \Cref{prop:rate_md}}\label{sec:proof_th_rate}

The proof below is written for a generic $L$-smooth and $l$-strongly convex functional $\cF$ relatively to a Bregman potential $\phi$. Recall that in the case of $\cF$ being the Kullback-Leibler divergence and negative entropy, $L=l=1$.\\

We first state a preliminary result, 
known as the "three-point inequality" or "Bregman proximal inequality", which can be found in \citep[Lemma 3]{aubin2022mirror}.

\begin{lemma}[Three-point inequality]\label{lem:three-point}
	Given $\mu\in\cM(\X)$ and some proper convex functional $\cG:\cM(\X)\rightarrow \R\cup\{+\infty\}$, if $\nabla\phi(\mu)$ exists, as well as $\bar{\nu}=\argmin_{\nu \in C}\{\cG(\nu)+B_{\phi}(\nu | \mu)\}$, then for all $\nu\in C \cap \dom(\phi) \cap \dom(\cG)$: \begin{equation}\label{eq:three_point_ineq}
		\hspace*{-0.2cm} \cG(\nu)+ B_{\phi}(\nu| \mu) \ge \cG(\bar{\nu}) + B_{\phi}(\bar{\nu}| \mu) +  B_{\phi}(\nu| \bar{\nu}).
	\end{equation}
\end{lemma}

We can now start the proof of mirror descent convergence rate.
	Since $\cF$ is $L$-smooth relative to $\phi$ and $\gamma_{n+1} <1/L$ implies, we have
 \begin{align}\label{eq:conv_smooth_new}
		\cF(\mu_{n+1})&\le \cF(\mu_n) + \ssp{\nabla \cF(\mu_n),\mu_{n+1}	-\mu_n}+L B_{\phi}(\mu_{n+1}|\mu_n)\\
  &\le \cF(\mu_n) + \ssp{\nabla \cF(\mu_n),\mu_{n+1}	-\mu_n}+\frac{1}{\gamma_{n+1}} B_{\phi}(\mu_{n+1}|\mu_n)
  .\nonumber
	\end{align}
	Applying Lemma \ref{lem:three-point} to the convex function $\cG_n(\nu)=\gamma_{n+1} \ssp{\nabla \cF(\mu_n),\nu-\mu_n}$, with $\mu=\mu_n$ and $\bar{\nu}=\mu_{n+1}$ yields
	\begin{equation*}
		\ssp{\nabla \cF(\mu_n),\mu_{n+1}	-\mu_n} + \frac{1}{\gamma_{n+1}} B_{\phi}(\mu_{n+1}|\mu_n) \le \ssp{\nabla \cF(\mu_n),\nu-\mu_n} + \frac{1}{\gamma_{n+1}} B_{\phi}(\nu| \mu_n) - \frac{1}{\gamma_{n+1}} B_{\phi}(\nu| \mu_{n+1}).
	\end{equation*}
	Fix $\nu$, then \eqref{eq:conv_smooth_new} becomes:\begin{equation}\label{eq:conv_3point_new}
		\cF(\mu_{n+1})\le \cF(\mu_n) + \ssp{\nabla \cF(\mu_n),\nu-\mu_n}	+ \frac{1}{\gamma_{n+1}}B_{\phi}(\nu| \mu_n) - \frac{1}{\gamma_{n+1}} B_{\phi}(\nu| \mu_{n+1}).
	\end{equation}
	This shows in particular, by substituting $\nu=\mu_n$ and since $B_{\phi}(\nu| \mu_{n+1})\ge 0$, that 
 \begin{align}
     \label{eq:kl_decreasing}
     \cF(\mu_{n+1})\le \cF(\mu_n)-\frac{1}{\gamma_{n+1}} B_{\phi}(\mu_n| \mu_{n+1}),
 \end{align}
i.e. $\cF$ is decreasing at each iteration. Since $\cF$ is $l$-strongly convex relative to $\phi$, we also have:
	\begin{equation*}
		\ssp{\nabla \cF(\mu_n),\nu-\mu_n}\le \cF(\nu)-\cF(\mu_n) - l B_{\phi}(\nu|\mu_n) 
	\end{equation*}
	and \eqref{eq:conv_3point_new} becomes:
\begin{equation}\label{eq:conv_strcvx_new}
		\cF(\mu_{n+1})\le \cF(\nu) + \left(\frac{1}{\gamma_{n+1}}-l\right) B_{\phi}(\nu| \mu_{n}) - \frac{1}{\gamma_{n+1}}B_{\phi}(\nu|\mu_{n+1}),
	\end{equation} 
 i.e., multiplying the previous equation by $(\gamma_{n+1}^{-1}-l)^{-1}$, we get
 \begin{equation}\label{eq:inequality_for_recursion}
	\left(\frac{1}{1- \gamma_{n+1} l}\right)	\cF(\mu_{n+1})\le \left( \frac{1}{1- \gamma_{n+1} l} \right)\cF(\nu) +\frac{1}{\gamma_{n+1}} B_{\phi}(\nu| \mu_{n}) - \frac{1}{\gamma_{n+1}}\left( \frac{1}{1- \gamma_{n+1} l} \right) B_{\phi}(\nu|\mu_{n+1}).
	\end{equation} 
	
\subsection{Proof for $(\gamma_n)_{n\ge 1}$ decreasing or constant}	\label{sec:decreasing_or_constant_step_size}
	Similarly to \citet{lu2018relatively}, we now consider for $n \ge 1$:
 \begin{equation*}
     \mathcal{P}(n): \quad \sum_{k=1}^{n} \left(\frac{1}{1-\gamma_k l}\right)^k \cF(\mu_{k})\le \sum_{k=1}^{n} \left(\frac{1}{1-\gamma_k l}\right)^k \cF(\nu) + \frac{1}{\gamma_1}  B_{\phi}(\nu| \mu_{0}) - \frac{1}{\gamma_n} \left(\frac{1}{1-\gamma_n l}\right)^n B_{\phi}(\nu|\mu_{n}).
 \end{equation*}
We first have that 
\begin{equation*}
    \mathcal{P}(1): \left(\frac{1}{1- \gamma_{1} l}\right)	\cF(\mu_{1})\le \left( \frac{1}{1- \gamma_{1} l} \right)\cF(\nu) +\frac{1}{\gamma_{1}} B_{\phi}(\nu| \mu_{0}) - \frac{1}{\gamma_{1}}\left( \frac{1}{1- \gamma_{1} l} \right) B_{\phi}(\nu|\mu_{1})
\end{equation*} is true by \eqref{eq:inequality_for_recursion}. Then, assume $\mathcal{P}(n)$ holds. We have by \eqref{eq:inequality_for_recursion}:
\begin{multline*}
    \sum_{k=1}^{n+1}  \left(\frac{1}{1-\gamma_k l}\right)^k \cF(\mu_{k}) = \sum_{k=1}^{n} \left(\frac{1}{1-\gamma_k l}\right)^k \cF(\mu_{k}) +\left(\frac{1}{1-\gamma_{n+1} l}\right)^{n+1} \cF(\mu_{n+1}) \\
    \le \sum_{k=1}^{n} \left(\frac{1}{1-\gamma_k l}\right)^k \cF(\nu) + \frac{1}{\gamma_1}  B_{\phi}(\nu| \mu_{0}) - \frac{1}{\gamma_n} \left(\frac{1}{1-\gamma_n l}\right)^n B_{\phi}(\nu|\mu_{n})\\ 
    + \left( \frac{1}{1- \gamma_{n+1} l} \right)^{n+1}\cF(\nu) +\frac{1}{\gamma_{n+1}}\left( \frac{1}{1- \gamma_{n+1} l} \right)^{n} B_{\phi}(\nu| \mu_{n}) - \frac{1}{\gamma_{n+1}}\left( \frac{1}{1- \gamma_{n+1} l} \right)^{n+1} B_{\phi}(\nu|\mu_{n+1})\\
    \le  \sum_{k=1}^{n+1} \left(\frac{1}{1-\gamma_k l}\right)^k \cF(\nu) + \frac{1}{\gamma_1}  B_{\phi}(\nu| \mu_{0})- \frac{1}{\gamma_{n+1}}\left( \frac{1}{1- \gamma_{n+1} l} \right)^{n+1} B_{\phi}(\nu|\mu_{n+1})
\end{multline*}
where we used in the last inequality (to upper bound the sum of the third and fifth term by zero) that $s\mapsto s^{-1}(1-s)^{-n}$ was a monotone increasing function and that the sequence $(\gamma_n)_{n\ge 1}$ was decreasing, showing $\mathcal{P}(n+1)$ holds. Hence $\mathcal{P}(n)$ is true for all $n\ge 1$. Then, using the monotonicity of $(\cF(\mu_n))_{n\ge 0}$ on the left hand side and the positivity of $B_{\phi}(\nu|\mu_n)$ on the right hand side of $\mathcal{P}(n)$, we have
	\begin{equation*}
	    \sum_{k=1}^{n} \left(\frac{1}{1-\gamma_k l}\right)^k \left(\cF(\mu_n) - \cF(\nu) \right)\le \frac{1}{\gamma_1}  B_{\phi}(\nu| \mu_{0})- \frac{1}{\gamma_n} \left(\frac{1}{1-\gamma_n l}\right)^n B_{\phi}(\nu|\mu_{n})\le \frac{1}{\gamma_1}  B_{\phi}(\nu| \mu_{0}).\qedhere
	\end{equation*}
This shows that 
\begin{equation*}
    \cF(\mu_n) - \cF(\mu) \le \frac{C_n}{\gamma_1}  B_{\phi}(\nu| \mu_{0}),\text{ where } C_n^{-1} = \sum_{k=1}^{n} \left(\frac{1}{1-\gamma_k l}\right)^k.
\end{equation*}

\subsection{Proof for general $(\gamma_n)_{n\ge 1}$}\label{sec:general_step_size}

Consider for $n \ge 1$:
 \begin{equation*}
     \mathcal{P}(n): \quad \sum_{k=1}^{n} \frac{\gamma_k}{\gamma_1}\prod_{i=1}^k\frac{1}{1-\gamma_i l} \cF(\mu_{k})\le \sum_{k=1}^{n}  \frac{\gamma_k}{\gamma_1}\prod_{i=1}^k\frac{1}{1-\gamma_i l} \cF(\nu) + \frac{1}{\gamma_1} B_{\phi}(\nu| \mu_{0}) -\frac{1}{\gamma_1}\prod_{i=1}^n\frac{1}{1-\gamma_i l} B_{\phi}(\nu|\mu_{n})
 \end{equation*}
We first have that 
\begin{equation*}
    \mathcal{P}(1): \left(\frac{1}{1- \gamma_{1} l}\right)	\cF(\mu_{1})\le \left( \frac{1}{1- \gamma_{1} l} \right)\cF(\nu) + \frac{1}{\gamma_1}B_{\phi}(\nu| \mu_{0}) - \frac{1}{\gamma_1}\left( \frac{1}{1- \gamma_{1} l} \right) B_{\phi}(\nu|\mu_{1})
\end{equation*} is true by \eqref{eq:inequality_for_recursion}. Then, assume $\mathcal{P}(n)$ holds. We have by \eqref{eq:inequality_for_recursion}:
\begin{multline*}
    \sum_{k=1}^{n+1} \frac{\gamma_k}{\gamma_1}\prod_{i=1}^k\frac{1}{1-\gamma_i l} \cF(\mu_{k}) = \sum_{k=1}^{n} \frac{\gamma_k}{\gamma_1}\prod_{i=1}^k\frac{1}{1-\gamma_i l} \cF(\mu_{k}) + \frac{\gamma_{n+1}}{\gamma_1}\prod_{i=1}^{n+1}\frac{1}{1-\gamma_i l}\cF(\mu_{n+1}) \\
    \le \sum_{k=1}^{n}  \frac{\gamma_k}{\gamma_1}\prod_{i=1}^k\frac{1}{1-\gamma_i l} \cF(\nu) +  \frac{1}{\gamma_1}B_{\phi}(\nu| \mu_{0}) -  \frac{1}{\gamma_1}\prod_{i=1}^n\frac{1}{1-\gamma_i l} B_{\phi}(\nu|\mu_{n}) \\
    + \frac{\gamma_{n+1}}{\gamma_1}\prod_{i=1}^{n}\frac{1}{1-\gamma_i l}\left(\left( \frac{1}{1- \gamma_{n+1} l} \right)\cF(\nu) +\frac{1}{\gamma_{n+1}} B_{\phi}(\nu| \mu_{n}) - \frac{1}{\gamma_{n+1}}\left( \frac{1}{1- \gamma_{n+1} l} \right) B_{\phi}(\nu|\mu_{n+1})\right)\\
    = \sum_{k=1}^{n+1}  \frac{\gamma_k}{\gamma_1}\prod_{i=1}^k\frac{1}{1-\gamma_i l} \cF(\nu) + \frac{1}{\gamma_1}  B_{\phi}(\nu| \mu_{0})  - \frac{1}{\gamma_1}\prod_{i=1}^{n+1}\frac{1}{1-\gamma_i l} B_{\phi}(\nu|\mu_{n+1}),
\end{multline*}
showing $\mathcal{P}(n+1)$ holds. Hence $\mathcal{P}(n)$ is true for all $n\ge 1$. Then, using the monotonicity of $(\cF(\mu_n))_{n\ge 0}$ on the left hand side and the positivity of $B_{\phi}(\nu|\mu_n)$ on the right hand side of $\mathcal{P}(n)$, we have
	\begin{equation*}
	    \sum_{k=1}^{n} \frac{\gamma_k}{\gamma_1}\prod_{i=1}^k\frac{1}{1-\gamma_i l}  \left(\cF(\mu_n) - \cF(\nu) \right)\le \frac{1}{\gamma_1}B_{\phi}(\nu| \mu_{0}) -\frac{1}{\gamma_1}  \prod_{i=1}^n\frac{1}{1-\gamma_i l} B_{\phi}(\nu|\mu_{n}) \le  \frac{1}{\gamma_1}B_{\phi}(\nu| \mu_{0}).\qedhere
	\end{equation*}
This shows that 
\begin{equation*}
    \cF(\mu_n) - \cF(\mu) \le \frac{C_n}{\gamma_1}  B_{\phi}(\nu| \mu_{0}),\text{ where } C_n^{-1} = \sum_{k=1}^n\frac{\gamma_k}{\gamma_1}\prod_{i=1}^k\frac{1}{1-\gamma_i l}.
\end{equation*}
In particular for $l=1$ and $\gamma_k = (\lambda_k -\lambda_{k-1})/(1-\lambda_{k-1}) $, 
\begin{equation*}
    C_n^{-1} = \frac{1}{\lambda_1}\sum_{k=1}^{n} \left(\frac{\lambda_k - \lambda_{k-1}}{1-\lambda_{k-1}}\frac{1-\lambda_0}{1-\lambda_k}\right) = \frac{1}{\lambda_1}\sum_{k=1}^{n} \left(\frac{\lambda_k - \lambda_{k-1}}{(1-\lambda_{k-1})(1-\lambda_k)}\right),
\end{equation*}
where we used $\lambda_0 = 0$.

\subsection{Bounds on convergence rate}

In this section we prove upper bounds on the convergence rates previously obtained. Our bounds are obtained in the case $l=1$ and $\gamma_n\leq 1/L=1$ for all $n\ge 1$.


We show that 
\begin{equation}
\label{eq:rate_bound1}
 C_n = \left(\sum_{k=1}^n\frac{\gamma_k}{\gamma_1}\prod_{i=1}^k\frac{1}{1-\gamma_i}\right)^{-1}\leq \prod_{k=1}^n(1-\gamma_k).
\end{equation}
To see this we consider for $n \ge 1$, $\mathcal{P}(n): \quad \sum_{k=1}^{n}\frac{\gamma_k}{\gamma_1}\prod_{i=1}^k\frac{1}{1-\gamma_i} \geq \prod_{k=1}^n(1-\gamma_k)^{-1}$.
We trivially have that  $\mathcal{P}(1): \left(\frac{1}{1-\gamma_1 }\right)^1 \geq (1-\gamma_1)^{-1}$ is true. Then, assume $\mathcal{P}(n)$ holds.
We have 
\begin{multline*}
\sum_{k=1}^{n+1} \frac{\gamma_k}{\gamma_1}\prod_{i=1}^k\frac{1}{1-\gamma_i} = \sum_{k=1}^{n} \frac{\gamma_k}{\gamma_1}\prod_{i=1}^k\frac{1}{1-\gamma_i} + \frac{\gamma_{n+1}}{\gamma_1}\prod_{i=1}^{n+1}\frac{1}{1-\gamma_i} \geq \prod_{k=1}^n(1-\gamma_k)^{-1} + \gamma_{n+1}\prod_{k=1}^{n+1}(1-\gamma_k)^{-1}\\
=\prod_{k=1}^n(1-\gamma_k)^{-1} \left[1+ \gamma_{n+1}(1-\gamma_{n+1})^{-1}\right] = \prod_{k=1}^{n+1}(1-\gamma_k)^{-1}
\end{multline*}
since $\gamma_1 \leq 1$ for all $n\ge 1$, showing $\mathcal{P}(n+1)$ holds. Hence \eqref{eq:rate_bound1} is true for all $n\ge 1$.

\section{Proof of Proposition \ref{prop:fdiv}}\label{sec:proof_fdiv}

\subsection{Tempering sequence as a parametric model}
Let us recall that the tempering sequence is defined as:
\[
\mu_{\lambda}(x)=\frac{\mu_0^{1-\lambda}(x)\pi^{\lambda}(x)}{\exp\left\{ \psi(\lambda)\right\} }=\mu_0(x)\exp\left\{ \lambda s(x)-\psi(\lambda)\right\} 
\]
for $\lambda\in[0,1]$, where $s(x):=\log \pi(x)/\mu_0(x)$, and 
\[
\psi(\lambda):=\log\int \mu_0(x)\exp\left\{ \lambda s(x)\right\} dx
\]
is the partition function (log-normalizing constant). 

In our case, the score is: $t_{\lambda}(x)=s(x)-\psi'(\lambda)$
(the score has expectation zero, as expected since $\psi'(\lambda)=\mathbb{E}_{\lambda}[s(X)]$),
and 
\[
I(\lambda):=\mathrm{Var}_{\lambda}\left[t_{\lambda}(X)\right]=\mathbb{E}_{\lambda}\left[t_{\lambda}(X)^{2}\right]=\mathrm{Var}_{\lambda}\left[s(X)\right].
\]

Note also the well-known identity: 
\begin{equation*}
I(\lambda)=-\mathbb{E}_{\lambda}[t'_{\lambda}(X)]=-\mathbb{E}_{\lambda}\left[\frac{\partial^{2}\log\mu_{\lambda}(X)}{\partial\lambda^{2}}\right].
\end{equation*}

\subsection{Proof}
Recall that $f$ must be convex and such that $f(1)=0$.

By the standard properties of $f-$divergence, it is clear that the
function $\varphi_{\lambda}:$ $\lambda'\rightarrow D_{f}(\lambda'|\lambda)$
is non-negative, and zero at $\lambda'=\lambda$, hence its first derivative
must be zero at $\lambda'=\lambda$. In fact, 
\begin{align*}
\varphi'_{\lambda}(\lambda') & =\int\mu_{\lambda'}'f'(\mu_{\lambda'}/\mu_{\lambda}) =\int\mu_{\lambda'}t_{\lambda'}f'(\mu_{\lambda'}/\mu_{\lambda})
\end{align*}
and note we have indeed $\varphi'_{\lambda}(\lambda)=f'(1)\int\mu_{\lambda}'=0$.
For the second derivative 
\[
\varphi''_{\lambda}(\lambda')=\int\mu_{\lambda'}(t_{\lambda'}'+t_{\lambda'}^{2})f'\left(\frac{\mu_{\lambda'}}{\mu_{\lambda}}\right)+\int\frac{\left(\mu_{\lambda'}t_{\lambda'}\right)^{2}}{\mu_{\lambda}}f''\left(\frac{\mu_{\lambda'}}{\mu_{\lambda}}\right)
\]
and at $\lambda'=\lambda$:
\begin{align*}
\varphi''_{\lambda}(\lambda) & =f'(1)\int\mu_{\lambda}(t_{\lambda}'+t_{\lambda}^{2})+f''(1)\int\mu_{\lambda}\left(t_{\lambda}\right)^{2}\\
 & =f''(1)I(\lambda).
\end{align*}
This ends the proof.

\section{Mirror descent is a time-discretization of Fisher-Rao flow}\label{sec:fisher_rao}

Mirror descent iteration on $\cF$ starting from $\mu_0$, is an Euler (or Forward) time-discretisation of the Fisher-Rao flow of $\cF$ \cite{lu2023birth}.
Indeed, the FR flow of a functional $\cF$ can be written
\begin{align*}
    \frac{\partial \mu_t}{\partial t} = - \mu_t \cF'(\mu_t),\text{ hence, } \frac{\partial \log(\mu_t)}{\partial t} 
    = 
    -\cF'(\mu_t). 
\end{align*}
An Euler discretization of the previous continuous dynamics write:
\begin{equation}
     \log(\mu_{l+1}) - \log(\mu_l) 
    = 
    - \gamma_{l+1} \cF'(\mu_l)
    \enspace,
\end{equation}
which recovers \eqref{eq:md} by exponentiating the equality.

\section{Algorithms details}\label{sec:algorithm_details}

We collect here further details on the SMC samplers described in \Cref{sec:algorithms} and  describe other strategies based on importance sampling that approximate the mirror descent iterates \eqref{eq:md}.

\begin{algorithm}[th]
\begin{algorithmic}[1]
\STATE{\textit{Inputs:} sequences of  temperatures $(\lambda_n)_{n= 1}^T$, Markov kernels $(M_n)_{n= 1}^T$, initial proposal $\mu_0$.}
\STATE{\textit{Initialize:} set $\lambda_0=0$, sample $\widetilde{X}_0^i\sim \mu_0$ and set $W_0^i=1/N$ for $i=1,\dots, N$.}
\FOR{$n=1,\dots, T$}
\IF{$n>1$}
\STATE{\textit{Resample:} draw $ \{\widetilde{X}_{n-1}^i\}_{i=1}^N$ independently from $\{X_{n-1}^i, W_{n-1}^i\}_{i=1}^N$ and set $W_n^i =1/N$ for $i=1,\dots, N$.}
\ENDIF
\STATE{\textit{Propose:} draw $X_n^i\sim M_{n}(\cdot, \widetilde{X}_{n-1}^i)$ for $i=1,\dots, N$.}
\STATE{\textit{Reweight:} compute and normalize the weights $W_n^i \propto w_{n}(\widetilde{X}_{n-1}^i)$ for $i=1,\dots, N$.}
\ENDFOR
\STATE{\textit{Output:} $q_n(x) =\sum_{i=1}^N W_n^i\delta_{X_n^i}(x)$}
\end{algorithmic}
\caption{SMC samplers \citep{del2006sequential}.}\label{alg:smc}
\end{algorithm}

\subsection{Other schemes}

\textbf{Particle Mirror Descent (PMD).} Similarly to SMC, \cite{dai2016provable} propose an approximation of the mirror descent iterates~\eqref{eq:md_temp} based on importance sampling. The mirror descent iterate at time $n$ is approximated by a kernel density estimator (KDE)
\begin{align}\label{eq:kde_pmd}
    q_n^{\textrm{PMD}}(x):=
    \sum_{i=1}^N V_n^i K_{h_n}(x - X_n^i),
\end{align}
where $\{X_n^i, V_n^i\}_{i=1}^N$ denotes a weighted particle set and $K_{h_n}$ is a smoothing kernel with bandwidth $h_n$.
At iteration $n$ the weighted particle set $\{X_{n-1}^i, V_{n-1}^i\}_{i=1}^N$ is resampled to obtain the equally weighted particle set $\{\widetilde{X}_{n-1}^i, 1/N\}_{i=1}^N$, the kernel $K_{h_n}$ is applied to propose new particle locations $X_n^i\sim K_{h_n}(\cdot - \widetilde{X}_{n-1}^i)$. The weights for the proposed particle set are then proportional to
\begin{align}\label{eq:pmd_weight}
v_n(x)=\left(\frac{\pi(x)}{q_{n-1}^{\textrm{PMD}}(x)}\right)^{\gamma_{n}}.
\end{align}
It is then clear that the Particle Mirror Descent (PMD) scheme summarized in \Cref{alg:pmd} in the Appendix is of the form \eqref{eq:approximate_scheme}.

Comparing PMD 
with SMC
and with the vast literature on SMC algorithms (see, e.g., \cite{chopin2020introduction} for a comprehensive introduction) we find that PMD
is an SMC algorithm targeting the sequence of distributions
\begin{align}
\label{eq:mu_tilde}
    \tilde{\mu}_{n}(x) \propto \int \mu_{n-1}(x')K_{h_n}(x-x')dx'\left(\frac{\pi(x)}{\eta_{n-1}(x)
    }\right)^{\gamma_{n}},
\end{align}
which converges to the mirror descent iterates~\eqref{eq:md_temp} as $h_n\to 0$.
The kernel $K_{h_n}$ is replacing the $\mu_n$-invariant kernel $M_n$ as proposal and the importance weights are given by~\eqref{eq:pmd_weight}.
However, PMD is not a standard SMC algorithm, since the weights $v_n$ are approximations of the idealized weights $v_n^\star(x) = \pi(x)/\int K_{h_n}(x-x')\mu_{n-1}(x')dx'$
obtained by plugging the KDE $q_{n-1}^{\textrm{PMD}}$ in place of the denominator. Hence PMD uses one more approximation than standard SMC samplers.

Leveraging the connection between mirror descent and tempering established in \Cref{sec:tempering}, it is easy to see that $v_n^\star\to w_n$ as $h_n\to 0$ (see~\eqref{eq:smc_pmd_weight}).
Hence, we could replace $v_n$ with $w_n$ in \Cref{alg:pmd} to reduce its computational cost and numerical error, since $v_n$ requires an $\mathcal{O}(N)$ cost due to the presence of the kernel density estimator $q_{n-1}$, while the cost of $w_n$ is $\mathcal{O}(1)$.
Nevertheless, this does not lead to an SMC algorithm targeting $\tilde{\mu}_{n}$ (or $\mu_n$).

\begin{algorithm}[th]
\begin{algorithmic}[1]
\STATE{\textit{Inputs:} sequences of  bandwidths $(h_n)_{n = 1,\ldots, T}$, learning rates $(\gamma_n)_{n=1, \ldots, T}$, initial proposal $\mu_0$.}
\STATE{\textit{Initialize:} sample $\widetilde{X}_0^i\sim \mu_0$ and set $W_0^i=1/N$ for $i=1,\dots, N$.}
\FOR{$n=1,\dots, T$}
\IF{$n>1$}
\STATE{\textit{Resample:} draw $ \{\widetilde{X}_{n-1}^i\}_{i=1}^N$ independently from $\{X_{n-1}^i, V_{n-1}^i\}_{i=1}^N$ and set $V_n^i =1/N$ for $i=1,\dots, N$.}
\ENDIF
\STATE{\textit{Propose:} draw $X_n^i\sim K_{h_n}(\cdot - \widetilde{X}_{n-1}^i)$ for $i=1,\dots, N$.}
\STATE{\textit{Reweight:} compute and normalize the weights $V_n^i \propto v_{n}(X_{n}^i)$ for $i=1,\dots, N$.}
\ENDFOR
\STATE{\textit{Output:} $q_{n}(x)= \sum_{i=1}^N V_n^i K_{h_n}(x - X_n^i)$.}
\end{algorithmic}
\caption{Particle Mirror Descent (PMD; \citet{dai2016provable}).}\label{alg:pmd}
\end{algorithm}


\textbf{Safe and Regularized Adaptive Importance Sampling (SRAIS)}.  \cite{korba2022adaptive}) propose an algorithm detailed in \Cref{alg:srais} in the Appendix, that samples at each iteration a particle $X_n$ from a proposal $q_{n}^{\textrm{SRAIS}}$.  Similarly to PMD which relies on the KDE estimator \eqref{eq:kde_pmd}, SRAIS relies on a KDE estimate to approximate the mirror descent iterates
\begin{equation}\label{eq:kde_srais}
    q_{n}^{\textrm{SRAIS}}(x)=\sum_{i=1}^{n} U_{i} K_{h_{i}}(x -X_i),
\end{equation}
where $\{X_i, U_i\}_{i=1}^n$ denotes a weighted particle set.
However, in this case the size of the particle population is not fixed and the KDE estimate uses all particles from previous iterations. Notice that the particle sampling step (Step 4 of \Cref{alg:srais}) can be repeated, resulting in sampling a batch $m_n$ of particles at step $n$. The weights for the proposed particles are
\begin{equation}
 u_n(x) =\left(\frac{\pi (x) }{ q_{n-1}^{\textrm{SRAIS}}(x)}\right)^{\gamma_{n}}\label{eq:srais_weights},
\end{equation}
and we can identify SRAIS to be of the form \eqref{eq:approximate_scheme}.
Similarly to PMD, one could replace $u_n$ with $w_n$ in SRAIS.
\begin{remark} In the original scheme proposed by \citet{korba2022adaptive}, the proposal at each iteration is defined as $\tilde{q}_{n+1}=(1-r_{n+1})q_{n+1}+r_{n+1} q_0$ where $q_{n+1}$ is the KDE~\eqref{eq:kde_srais}, $(r_n)_{n\ge 0}$ is a sequence in $[0,1]$ converging to $0$ 
and $q_0$ is a "safe" density (e.g. with heavy tails) preventing the importance weights from degeneracy. In \Cref{alg:srais} we removed the dependency with the safe density and took the sequence $(r_n)_{n\ge 0}$ constant equal to zero for a clearer presentation. 
\end{remark}

\begin{algorithm}
\begin{algorithmic}[1]
\STATE{\textit{Inputs:} Sequences of  bandwidths $(h_n)_{n = 1,\ldots, T}$, learning rates $(\gamma_n)_{n=1, \ldots, T}$, initial proposal $\mu_0$. }
\STATE{\textit{Initialize:} sample $X_1\sim \mu_0$ and set $U_1=(\pi(X_1)/\mu_0(X_1))^{\gamma_1}$.}
\FOR{$n= 1,\ldots, T $}
\STATE{\textit{Propose:}  draw $X_{n+1}\sim q_{n}$}
 \STATE{\textit{Reweight:} compute the weight $U_{n+1} \propto u_{n+1}(X_{n+1})$ and normalize the weights.}
\STATE{Update the proposal as in~\eqref{eq:kde_srais}.}
\ENDFOR
\STATE{\textit{Output:} $q_{n+1}(x)= \sum_{i=1}^{n+1} U^i K_{h_i}(x - X_i)$.}
 \end{algorithmic}
 \caption{Safe and Regularized Adaptive Importance sampling (SRAIS; \cite{korba2022adaptive})} \label{alg:srais}
\end{algorithm}

\subsection{Comparison of algorithms}

As discussed in the previous sections, both SMC samplers and PMD are an instance of SMC algorithms (albeit not a standard one in the case of PMD). 
The convergence properties of SMC defined in \Cref{alg:smc} 
are guaranteed by the wide literature on SMC algorithms (see, e.g., \cite{smc:theory:Del04} for a complete account). In particular, one can show (see \citet[Theorem 7.4.3]{smc:theory:Del04} and \cite{smc:theory:CD02}) that every measurable bounded function $\varphi:\mathbb{R}^d\to\mathbb{R}$ with $\Vert\varphi\Vert:=\sup_{x\in \mathbb{R}^d}\vert\varphi(x)\vert<\infty$,
\begin{align*}
      \mathbb{E}\left[\left\lvert \int \varphi \diff\mu_n-\int \varphi \diff q_n^{\textrm{SMC}}\right\rvert\right] &\leq \frac{B_n^{\textrm{SMC}}\Vert \varphi\Vert}{N^{1/2}}
\end{align*}
where $B_n^{\textrm{SMC}}$ denotes a finite constant which does not depend on $N$.

A similar result for \Cref{alg:pmd} has been established in \citet[Theorem 5]{dai2016provable}. The approximation error of PMD is divided into an optimisation error, due to the fact that the algorithm is stopped at time $T$, and the following approximation error arising from the particle approximation to the target $\tilde{\mu}_n$ in~\eqref{eq:mu_tilde}
\begin{align*}
    \mathbb{E}\left[\left\lvert \int \varphi \diff\tilde{\mu}_n -\int \varphi \diff q_n^{\textrm{PMD}}\right\rvert\right] &\leq \frac{B_n^{\textrm{PMD}}\Vert \varphi\Vert}{N^{1/2}},
\end{align*}
where $B_n^{\textrm{SMC}}$ denotes a finite constant which does not depend on $N$.

In the case of SMC samplers, there is no optimisation error since Algorithm~\ref{alg:smc} targets $\mu_n$ directly (and not the smoothed version~\eqref{eq:mu_tilde}) and, by construction, at time $T$ we have $\lambda_T =1$ so that $\mu_T = \pi$.

Furthermore, when implementing Algorithm~\ref{alg:smc} there is no need to introduce the kernel $K_{h_n}$ to obtain a KDE at each iteration, this results in a simpler algorithm than Algorithm~\ref{alg:pmd} which does not require the bandwidth parameter $h_n$ whose tuning is notoriously difficult \citep{silverman1986density}. Additionally, KDE performs poorly if the dimension of the underlying space is large \citep{chacon2018multivariate}.

The presence of the KDE in PMD also causes the algorithm to have a higher computational cost than standard SMC samplers, in fact, the presence of the KDE in the weights~\eqref{eq:pmd_weight} means that these weights require an $\mathcal{O}(N)$ cost to be computed  for each particle, against the $\mathcal{O}(1)$ per particle of the weights~\eqref{eq:smc_weight}. These results in a $\mathcal{O}(NT)$ cost for Algorithm~\ref{alg:smc} and $\mathcal{O}(N^2T)$ for Algorithm~\ref{alg:pmd}.
Clearly, the $\mathcal{O}(N^2T)$ of PMD could be reduced to $\mathcal{O}(NT)$ by replacing the weights~\eqref{eq:pmd_weight} with~\eqref{eq:smc_weight}, since the former are an approximation of the idealized weights $v_n^\star(x) = (\pi(x)/\mu_{n-1}(x))^{\gamma_{n}}$ which are proportional to~\eqref{eq:smc_weight} as shown in~\eqref{eq:smc_pmd_weight}, at the cost of targeting a slightly different distribution.

The computational cost of iteration $n$ of SRAIS is $\mathcal{O}(n)$ because of the KDE in the weights~\eqref{eq:srais_weights}. Hence, the cost of \Cref{alg:srais} is $\sum_{n=1}^T \mathcal{O}(n)\approx \mathcal{O}(T^2)$. In practice, to reduce computational cost, one could use only the last iterations as the first ones can be considered as ``burn-in'' steps.

\section{Further discussion on~\eqref{eq:ode_other_paper} and implementation details}
\label{app:odes}
Consider the well-known identity \citet[Section 4.4]{brekelmans2020all}
\begin{align}\label{eq:kl_integration}
    \KL(\mu_{\lambda_{n-1}}|\mu_{\lambda_n}) &= 
    \int_{\lambda_{n-1}}^{\lambda_n}(\lambda_{n}-\lambda)\mathrm{Var}_{\lambda}\left[s(X)\right] d\lambda\\& = \int_{\lambda_{n-1}}^{\lambda_n}(\lambda_n-\lambda)\FI(\lambda) d\lambda.\nonumber
\end{align}
We want to study the infinitesimal behaviour of the $\KL$ when $\lambda_{n-1} = \lambda(t)$ but $\lambda_n$ is fixed.
As suggested in \cite{goshtasbpour2023adaptive} a natural requirement is to keep the derivative of the $\KL$ w.r.t. time constant
\begin{align*}
    \frac{d}{dt}\KL(\mu_{\lambda(t)}|\mu_{\lambda_n}) &= 
    \frac{d}{dt}\left(\int_{\lambda(t)}^{\lambda_n}(\lambda_n-\lambda)\FI(\lambda) d\lambda\right)\\
    &=\frac{d\lambda(t)}{dt}(\lambda_n-\lambda(t))\FI(\lambda(t)) = c,
\end{align*}
where we used Leibniz integral rule for differentiation under the integral sign under the assumptions that all quantities are well-defined.
This gives us the following ODE for $\lambda(t)$
\begin{align}
\label{eq:ODE_appendix}
    \frac{d\lambda(t)}{dt} = c\left[(\lambda_n-\lambda(t))\FI(\lambda(t))\right]^{-1}.
\end{align}

If we set $\lambda_n=1$, i.e. we want to decrease the $\KL$ between $\mu_{\lambda(t)}$ and $\pi$ at a constant rate we obtain
\begin{align*}
    \frac{d\lambda(t)}{dt} = c\left[(1-\lambda(t))\FI(\lambda(t))\right]^{-1},
\end{align*}
i.e. the ODE given in \cite{goshtasbpour2023adaptive}, where we used the fact that 
\begin{align}
\label{eq:more_stable}
    \textrm{Var}_{\mu_{\lambda(t)}}\left(\log(\frac{\pi}{\mu_{\lambda(t)}}(x))\right) 
    = (1-\lambda(t))^2\textrm{Var}_{\mu_{\lambda(t)}}\left(\log(\frac{\pi}{\mu_0}(x))\right)=(1-\lambda(t))^2\FI(\lambda(t)).
\end{align}

If we instead assume that $\lambda_n$ is sufficiently close to $\lambda(t)$ that $\lambda_n-\lambda(t) \approx \nicefrac{d\lambda(t)}{dt}$ we obtain the ODE in~\eqref{eq:ode}. Or, equivalently, by discretizing~\eqref{eq:ODE_appendix} we obtain
\begin{align*}
    \lambda_n - \lambda_{n-1} = c\left[(\lambda_n-\lambda_{n-1})\FI(\lambda_{n-1})\right]^{-1},
\end{align*}
which is equivalent to~\eqref{eq:lambda_sequence}.

\subsection{Numerical implementation for Figure~\ref{fig:smc_vs_ais}}
\label{app:numerical_details}
We reproduce the narrow Gaussian experiment of \cite{goshtasbpour2023adaptive}: the target is $\pi = \mathcal{N}(1_d, 0.1^2\Id)$ and $\mu_0 = \mathcal{N}(0_d, \Id)$ where $d=2$.

To place both algorithms on equal footing we use the same number of particles $N=10^4$ and the same Markov kernels, i.e. random-walk Metropolis kernels automatically calibrated on the current particle sample.
In the case of SMC, we select the next tempering sequence so that $\ess_n = N/2$, or, equivalently, by setting $\beta =1$ in \eqref{eq:kl_decrease}.
For the constant rate AIS of \cite{goshtasbpour2023adaptive}, we follow their recommendation and set $\delta = 1/32$ (higher values of $\delta$ give slightly shorter tempering sequences but considerably worse approximations of $\pi$). To make their algorithm more numerically stable we replace line 11 in their Algorithm 1 with \eqref{eq:more_stable}. We point out that the resampling cost in SMC is negligible, and that a shorter tempering sequence does correspond to a shorter runtime ($<1$ second for SMC and $\approx 14$ seconds for AIS).

\end{document}